\documentclass[sigchi]{acmart}

\AtBeginDocument{%
  \RequirePackage{hyperxmp}
}
\pdfoutput=1 
\AtBeginDocument{%
  \providecommand\BibTeX{{%
    \normalfont B\kern-0.5em{\scshape i\kern-0.25em b}\kern-0.8em\TeX}}}

\setcopyright{acmcopyright}
\copyrightyear{2025}
\acmYear{2025}
\acmDOI{XXXXXXX.XXXXXXX}

\acmConference[Conference acronym 'XX]{Make sure to enter the correct
  conference title from your rights confirmation emai}{June 03--05,
  2018}{Woodstock, NY}
\acmBooktitle{Woodstock '18: ACM Symposium on Neural Gaze Detection,
 June 03--05, 2018, Woodstock, NY} 
\acmPrice{15.00}
\acmISBN{978-1-4503-XXXX-X/18/06}
\usepackage{tabularx,booktabs,array}
\usepackage[table]{xcolor} 
\newcolumntype{L}[1]{>{\raggedright\arraybackslash}p{#1}}
\newcolumntype{C}[1]{>{\centering\arraybackslash}p{#1}}

\usepackage{xcolor}

\newif\ifsubmit
\submitfalse
\ifsubmit
\newcommand{\hari}[1]{}
\newcommand{\sean}[1]{}
\newcommand{\colleen}[1]{}
\newcommand{\maneesh}[1]{}
\else
\newcommand{\hari}[1]{{\textcolor{purple}{\bf [*** HS: #1]}}}
\newcommand{\sean}[1]{{\textcolor{red}{\bf [*** SF: #1]}}}
\newcommand{\colleen}[1]{{\textcolor{blue}{\bf [*** CS: #1]}}}
\newcommand{\maneesh}[1]{{\textcolor{green}{\bf [*** MA: #1]}}}
\fi

\definecolor{SituationColor}{HTML}{5C8FA8}
\definecolor{ProcessColor}{HTML}{8C79B8}
\definecolor{FrameColor}{HTML}{C58B32}

\definecolor{SituationDark}{HTML}{3F677A}
\definecolor{ProcessDark}{HTML}{66588C}
\definecolor{FrameDark}{HTML}{8A6123}

\colorlet{SituationLight}{SituationColor!15}
\colorlet{ProcessLight}{ProcessColor!15}
\colorlet{FrameLight}{FrameColor!16}

\definecolor{SituationColor}{HTML}{3F86A8}   
\definecolor{ProcessColor}{HTML}{7656A6}     
\definecolor{FrameColor}{HTML}{C17A2B}       

\colorlet{SituationLight}{SituationColor!14}
\colorlet{ProcessLight}{ProcessColor!14}
\colorlet{FrameLight}{FrameColor!16}

\newcommand{\situationterm}[1]{\textcolor{SituationColor}{\textbf{#1}}}
\newcommand{\processterm}[1]{\textcolor{ProcessColor}{\textbf{#1}}}
\newcommand{\frameterm}[1]{\textcolor{FrameColor}{\textbf{#1}}}

\usepackage{amsthm}
\usepackage{enumitem}
\theoremstyle{definition}
\newtheorem{definition}{Definition}

\setlist[description]{leftmargin=1.8em, labelindent=1em, style=nextline}

\usepackage{comment}

\usepackage{url}
\usepackage{hyperref}

\begin{document}

\title[Alignment Games]{Alignment Games: A Framework for Conceptual Repair in Human–AI Collaboration}


\author{Hari Subramonyam}
 \affiliation{%
  \institution{Stanford University}
   \country{USA}
 }
 \email{harihars@stanford.edu}


 \author{Maneesh Agrawala}
 \affiliation{%
  \institution{Stanford University}
   \country{USA}
 }
 \email{magrawala@stanford.edu}

  \author{Sean Follmer}
 \affiliation{%
  \institution{Stanford University }
   \country{USA}
 }
 \email{sfollmer@stanford.edu}

\renewcommand{\shortauthors}{Subramonyam, et al.}

\begin{abstract}
The meaning of a concept in use is shaped by the situation, task, goals, and prior knowledge. For example, a request to make a poster ``visually appealing for a five-year-old'' might evoke bright colors and cartoon imagery for one collaborator, but less text, bold shapes, and visual simplicity for another.
We call such task-relevant differences \textit{conceptual misalignment}. We introduce \textit{Alignment Games}, a framework for making these differences visible and repairable during human--AI interaction. Drawing on theories of situated conceptualization, we characterize task-specific conceptual frames in terms of relevant attributes, values, relations, constraints, and priorities. We then define alignment moves that intervene on the situation, the reasoning used to interpret it, or the resulting frame. Through examples from educational content generation, creative coding, and argumentative writing, we show how these moves can be composed into repair sequences and derive design principles for supporting task-sufficient conceptual alignment at runtime.

\end{abstract}

\begin{teaserfigure}
  \includegraphics[width=\textwidth]{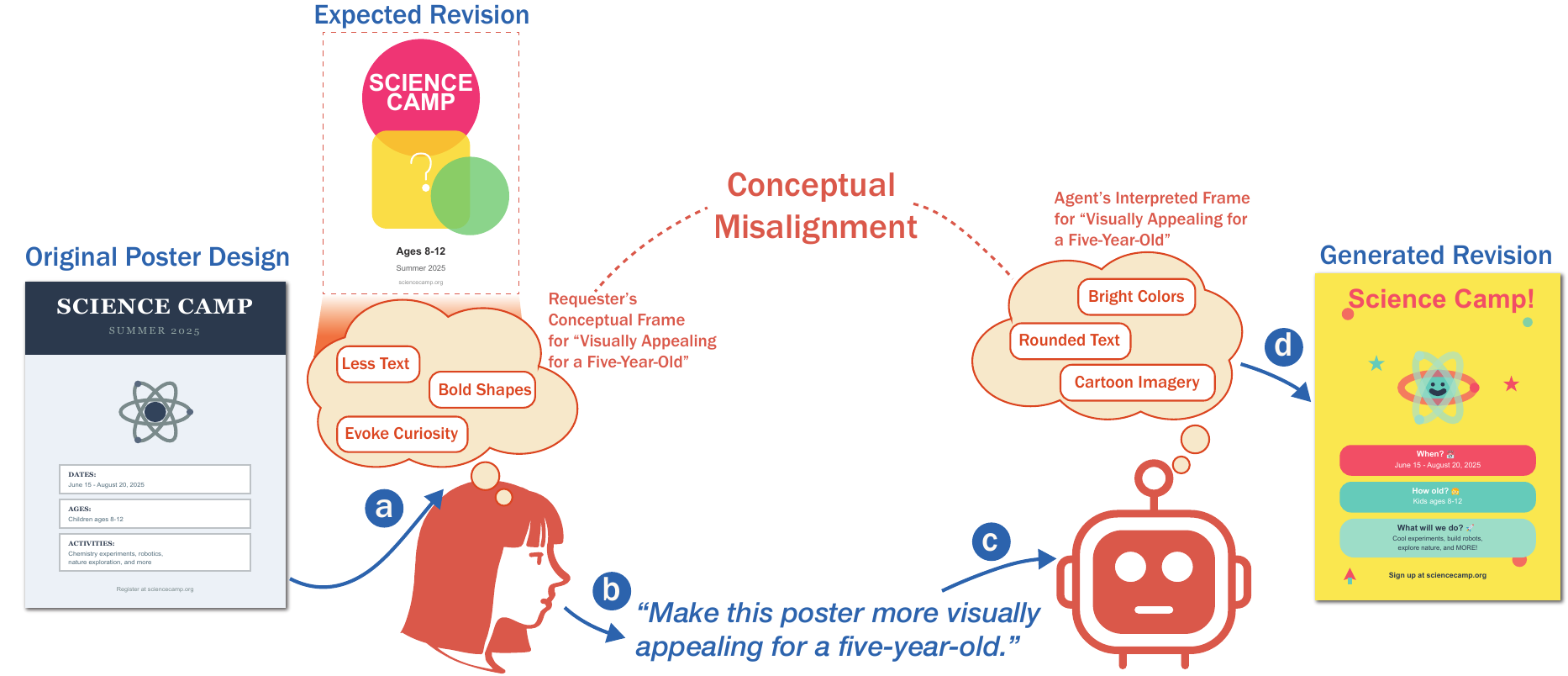}
  \caption{Conceptual Misalignment in Human-AI Collaboration: Given the same request to make a poster ``more visually appealing for a five-year-old,'' the requester (a) and AI agent (c) construct different conceptual frames from the compressed natural language instruction (b). The requester envisions bold shapes and minimal text (top), while the agent generates bright colors and cartoon imagery (right, d). Both interpretations are valid yet misaligned, demonstrating the challenge of conceptual coordination without explicit repair mechanisms.}
  \Description{
  Diagram of a requester and an AI agent each forming a different mental picture from the same instruction. A plain blue science-camp poster on the left branches into two revisions: the requester's imagined version (bold shapes, minimal text) at top, and the agent's generated version (bright yellow, cartoon atom, colorful buttons) at right.
  }
  \label{fig:teaser}
\end{teaserfigure}


\maketitle

\section{Introduction}
\label{sec:introduction}

Imagine being given the `original poster' in Figure~\ref{fig:teaser} and asked to \textit{``make it more visually appealing for a five-year-old.''} You might interpret this request as using bright colors, rounded typography, and cartoon imagery (Figure~\ref{fig:teaser}c). The requester, however, might have imagined less text, bolder shapes, and a design that evokes curiosity (Figure~\ref{fig:teaser}a). Neither interpretation is necessarily wrong and can plausibly satisfy the words in the request. However, such differences arise because concepts are not always fixed definitions transmitted intact through language and interpreted by retrieving those definitions from memory. Instead, people construct \textit{situated conceptualizations} that are context-sensitive representations constructed from prior experience that bring together the features, actions, goals, and other knowledge relevant to the situation at hand~\cite{barsalou1999perceptual,barsalou2008grounded}.

We use the term \emph{conceptual misalignment} for such task-relevant differences between collaborators' situated conceptualizations. In human-human collaborations, these differences are common and routinely repaired through interactions~\cite{pickering2021understanding}. Suppose the designer shows the requester a colorful revision filled with cartoon illustrations. The requester might respond, \textit{``The cartoons aren't really what I meant. I want it to feel simpler---something a child can understand from across the room.''} The designer might ask whether the problem is the imagery or the amount of information, show two alternatives, or suggest treating the poster more like a children's museum display than a birthday invitation. Through such examples, contrasts, clarifications, and reframing, the collaborators progressively learn not only \emph{what output} the other prefers, but \emph{how the other is construing the concept} that guides the work. This kind of repair is related to (but not identical with) establishing \emph{common ground}~\cite{clark1996using}. In the poster example, the collaborators may already share substantial common ground about the task such as they know which poster is under discussion, who its audience is, and what
request was made. Where their common ground is incomplete is in how they
conceptualize ``visually appealing for a five-year-old,'' i.e.,  which qualities are relevant, what values those qualities should take, how they relate, and which should be prioritized. 

This problem is especially consequential in interactions with generative AI. Users routinely ask systems to make an artifact ``clearer,'' a paragraph ``more optimistic,'' or a design ``more engaging,'' yet the system's operative \emph{interpretation} of such concepts typically remains implicit. When an output does not meet the expectation, the user must infer what the system understood differently, decide what to correct, and express that repair indirectly through another prompt or edit. This makes conceptual repair costly and uncertain. For instance, if a poster feels ``too childish,'' the mismatch might lie in its colors, typography, imagery, text density, or broader framing of the audience. A prompt such as ``make it less childish'' may fix one dimension, leave the underlying mismatch untouched, or alter aspects that were already satisfactory. Consequently, interaction can become an output-level search over possible repairs rather than a direct negotiation over the \textbf{meaning guiding generation}.

We introduce \emph{Alignment Games} (i.e., structured interactions in which
collaborators make moves and receive feedback to surface and repair conceptual misalignment ) as a framework for making this process of conceptual repair explicit and designable. Rather than only correcting the resulting artifact, collaborators can act on the interpretation that produced it. In the poster example, the system might surface that it is interpreting ``appealing for a five-year-old'' through attributes such as \emph{color}, \emph{typography}, and \emph{imagery}, with values such as \emph{bright}, \emph{rounded}, and \emph{cartoon-like}. The requester could then identify that \emph{text density} and \emph{visual simplicity} are missing, clarify that ``playful'' does not mean visually busy, prioritize readability over novelty, or reframe the design as a children's museum poster rather than a birthday invitation. We call these interventions \emph{alignment moves}; structured sequences of such moves form \emph{Alignment Games} through which collaborators diagnose, negotiate, and repair conceptual misalignment. 

To make these repairs systematic, we develop a \emph{process model of
conceptual alignment in human--AI interaction}, grounded in cognitive science. The model characterizes how external situations, internal task context, and prior knowledge interact through perception, retrieval, reasoning, and simulation to construct a situated conceptual frame. Differences between collaborators can therefore arise in what attributes they represent, the values assigned to them, how those attributes relate or are constrained, and which considerations are prioritized. This account provides the representational basis for describing both conceptual misalignment and the moves available for repairing it. Our contributions are (1) a \textbf{process model of situated conceptualization and conceptual misalignment} that characterizes how task-relevant interpretations are constructed and when differences become consequential; and (2) the \textbf{Alignment Games framework}, a grammar of moves for diagnosing, negotiating, and repairing conceptual misalignment across situational inputs, conceptual frames, and frame-construction processes.

\section{A Note on Framework Development}
\begin{figure*}
    \centering
    \includegraphics[width=\linewidth]{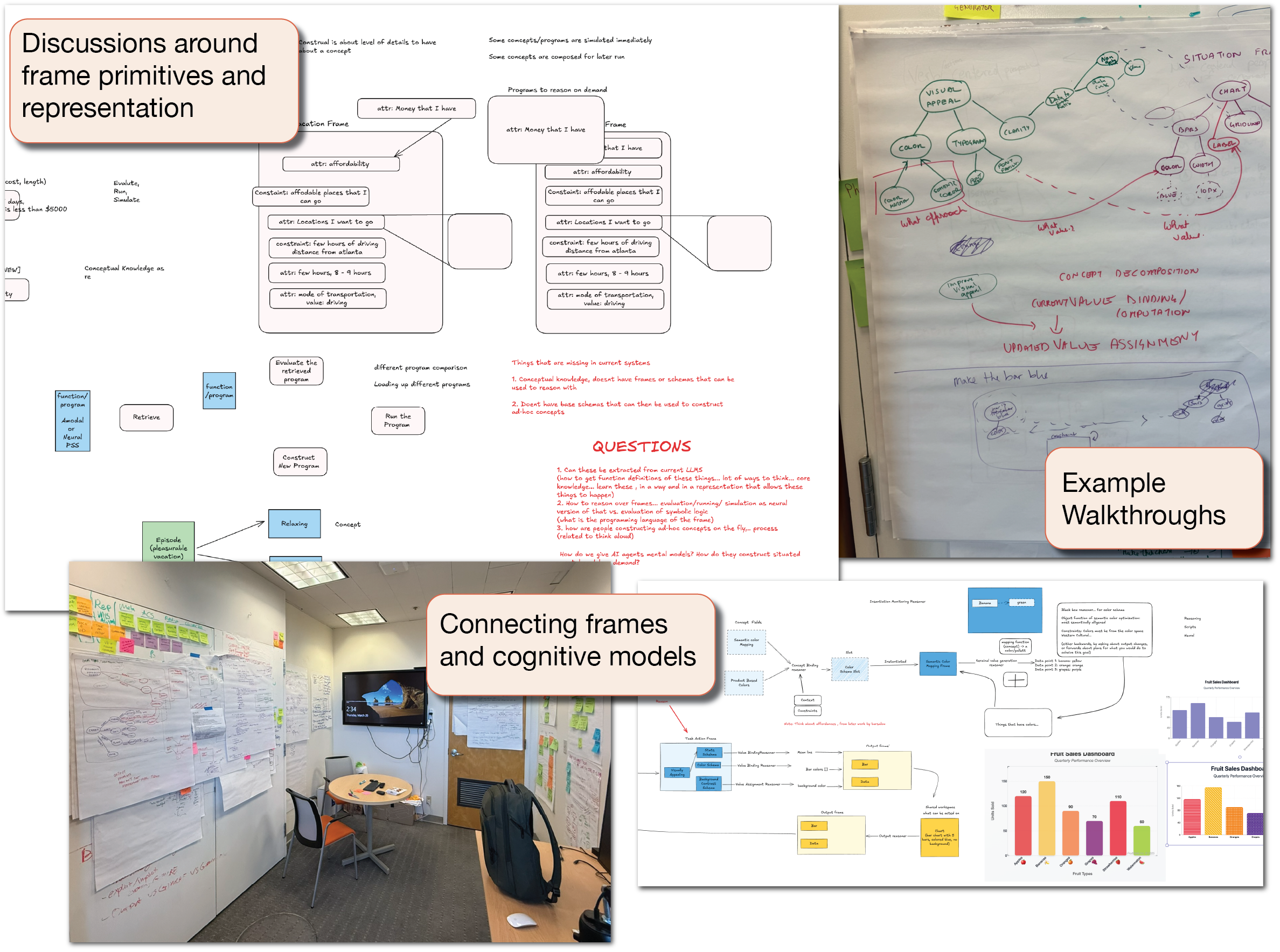}
    \caption{
   Examples from the iterative development of the Alignment Games framework. Across collaborative working sessions, we explored alternative frame primitives and representations, connected candidate representations to theories of situated cognition, and worked through concrete human--AI interaction examples to refine the framework and alignment-move vocabulary.
    }
    \Description{Collage of four photographs and screenshots from working sessions, each with a rounded label. Clockwise from top left: a digital whiteboard of boxed frame primitives and notes; handwritten whiteboard sketches of node-and-arrow concept maps; a flowchart mockup ending in bar-chart dashboards; and a meeting room with walls covered in sticky notes and paper diagrams.}
    \label{fig:work}
\end{figure*}

We developed Alignment Games through an iterative theory-building process combining literature synthesis, conceptual modeling, and analysis of existing human--AI interfaces. Across more than 200 hours of face-to-face collaborative analysis and discussion between the first and last author and weekly meetings with all authors, we generated and compared alternative representations of conceptual alignment, including different decompositions of conceptual frames, sources of misalignment, and candidate repair moves (Figure~\ref{fig:work}). We repeatedly tested these representations against breakdowns and interaction mechanisms reported in prior HCI systems, refining constructs when they collapsed distinct phenomena, failed to characterize a plausible repair, or introduced distinctions that did not appear useful for interaction design. This process converged on a task-relative situated conceptual frame and three loci for intervention---the situation, the processes used to construct an interpretation, and the resulting frame---from which we derived the alignment move vocabulary presented in Section~\ref{sec:alignment-games}.

\section{Situated Conceptualization and Conceptual Misalignment}
\label{sec:situated-conceptualization}

To characterize conceptual misalignment, we first need to understand how collaborators construct meaning for a particular task. Drawing on cognitive accounts of situated and grounded cognition, we develop a process model that addresses five questions: (1) What shapes a situated conceptualization? (2) How is it constructed? (3) What representation does it produce? (4) When are collaborators conceptually misaligned? (5) And when is their alignment sufficient for the task? Our goal is not to provide a complete cognitive architecture, but to identify the inputs, processes, and representational structures most relevant to understanding and repairing conceptual differences during collaboration.

\subsection{What Shapes a Situated Conceptualization?}
\label{sec:conceptual-inputs}

Broadly, a person's interpretation of a concept depends on information from three key sources: (1) the external situation, (2) their internal (mental) understanding of the task, and (3) their prior knowledge~\cite{barsalou1999perceptual,hinds1992context,bradley2005toward,reis2008reinvigorating,BARNETT2025101593}. \situationterm{External situational elements} are the features of the encountered task environment that provides perceptual and communicative grounding for conceptualization~\cite{reis2008reinvigorating}. In the poster example, these include the current poster, its visual organization, the science-camp context, available design tools, and the client's request. Situation models characterize such contexts along dimensions including space, time, entities, causality, and intentionality~\cite{zwaan1998situation}. These elements establish a frame of reference for interpretation by shaping what a person is likely to attend to, what prior knowledge becomes relevant, and what actions are possible~\cite{reis2008reinvigorating,barsalou1999perceptual,bradley2005toward}.

\situationterm{Internal task context} refers to the individual's current goals, values, motivations, role obligations, and resource constraints~\cite{locke2002building,ratneshwar2001goal}. A designer who prioritizes clarity may interpret visual appeal differently from one who prioritizes novelty. Likewise, a brand obligation or limited production time may change which possibilities are treated as viable. Internal task context filters both perception and retrieval, influencing which aspects of the situation become salient and which interpretations are preferred. Finally, people draw on \situationterm{prior knowledge and experience} in long-term memory. This includes schemas for recurring situations~\cite{bartlett1995remembering,rumelhart2017schemata}, scripts and plans for routine event sequences~\cite{schank2013scripts}, prototypes and exemplars~\cite{rosch1975cognitive,medin1978context}, causal or theory-based knowledge~\cite{murphy1985role}, and episodic memories of particular experiences~\cite{tulving1972episodic}. A designer may, for example, retrieve children's books, toy packaging, classroom materials, or previous design projects when interpreting what ``appealing for a five-year-old'' might entail. Because collaborators differ in experience and expertise, the same situational cue need not activate the same knowledge. Together, these sources provide the material from which a situated interpretation is constructed.

\subsection{How Is a Situated Conceptualization Constructed?}
\label{sec:model}

\begin{figure*}
    \centering
    \includegraphics[width=\linewidth]{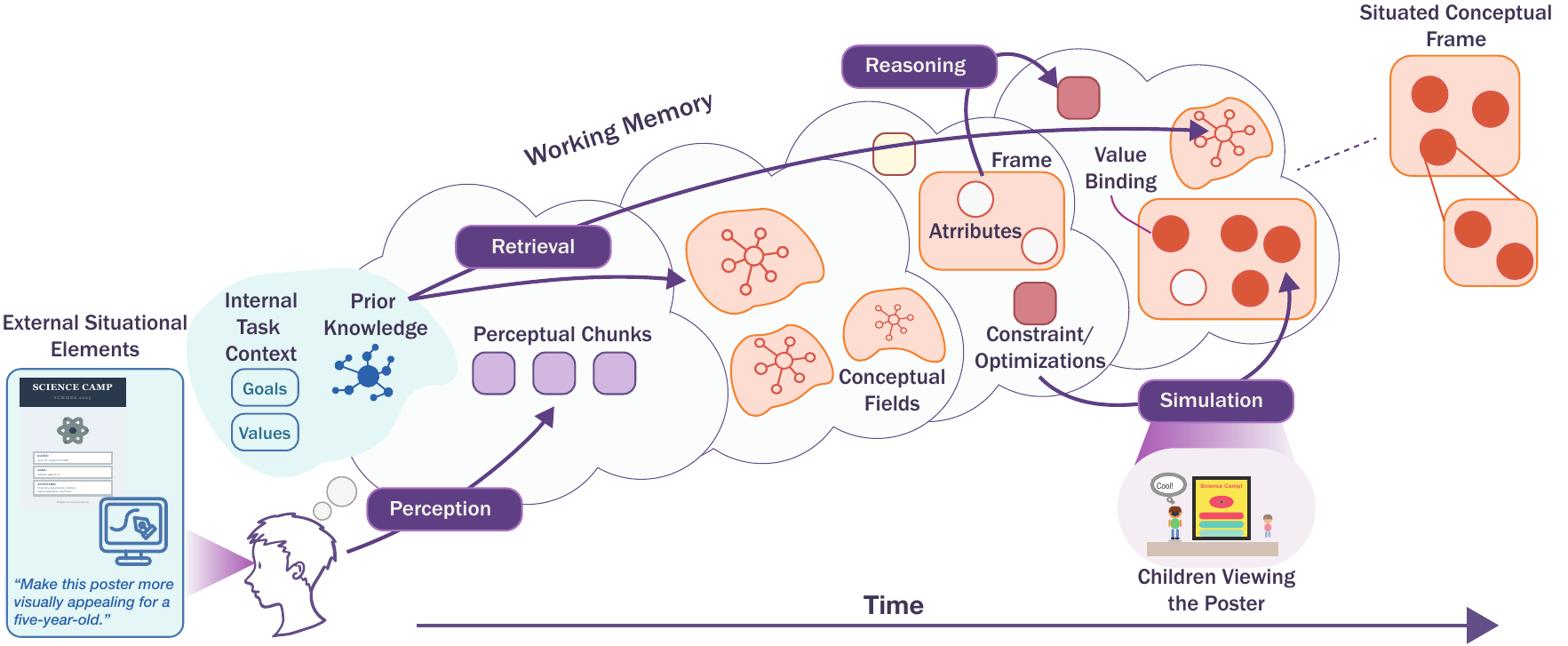}
    \caption{
    \textbf{Process model of situated conceptualization.}
    \textcolor{SituationDark}{External situational elements, internal task context, and prior knowledge}
    shape an iterative cycle of
    \textcolor{ProcessDark}{perception and attention, retrieval, reasoning, and simulation}
    in working memory. Through this cycle, a collaborator constructs a
    \textcolor{FrameDark}{situated conceptual frame}
    that organizes the attributes, values, relations, constraints, priorities, and broader framings relevant to the task. Actions guided by the frame alter the artifact or situation, producing new evidence that can trigger subsequent cycles of construction and revision.
    }
    \Description{
    Process diagram reading left to right along a "Time" arrow. A person views a science-camp poster and instruction; arrows labeled Perception, Retrieval, Reasoning, and Simulation move through a large cloud labeled Working Memory, where perceptual chunks and conceptual fields combine into a frame with attributes, values, and constraints, producing a Situated Conceptual Frame at the far right.
    }
    \label{fig:model}
\end{figure*}

We propose that the above sources interact through an iterative cycle of \processterm{perception and attention}, \processterm{retrieval}, \processterm{reasoning}, and \processterm{simulation} in working memory (Figure~\ref{fig:model}). 

\subsubsection{Perception} Construction of concept understanding begins by selectively recruiting information from the current situation. Bottom-up perceptual processes encode observable features, while top-down influences from goals and prior knowledge shape which features receive attention and how they are organized. In the poster example, muted colors, dense text, and formal typography may initially appear simply as visual properties. Relative to the goal of appealing to a young child, however, these same features may become deficiencies or opportunities for change. If the request were instead to ``make the poster clearer,'' typography and layout might become salient while muted colors remain acceptable. Thus, task context shapes not only how features are evaluated, but which features enter the conceptualization in the first place~\cite{barsalou1999perceptual,ratneshwar2001goal}.

\subsubsection{Retrieval} Perceived and attended features cue relevant knowledge from memory. A formal typeface may activate prior examples of children's typography; the science-camp context may activate concepts associated with experimentation, discovery, or play. Retrieval can introduce candidate attributes and values, exemplars, episodic experiences, and partially structured prior frames. Internal task context further constrains this process: a designer who values novelty may retrieve different associations than one prioritizing familiarity. Prior knowledge supplies raw material that can be selectively recruited and reorganized for the current task rather than simply providing a fixed concept.

\subsubsection{Reasoning} Retrieved information must then be organized into a coherent interpretation. Reasoning establishes which attributes matter, what values they should take, how they relate, and what constraints or priorities should govern them. In the poster example, the goal of visual appeal might establish preferred directions such as greater playfulness or stronger visual hierarchy, while the need for scientific accuracy or legibility constrains acceptable solutions. Reasoning can also reorganize conceptual structure through operations such as analogy, decomposition, abstraction, or conceptual blending. A designer might, for example, borrow the lively hierarchy of a birthday invitation while replacing balloons with science-related imagery.

\subsubsection{Simulation} People can additionally evaluate and extend an emerging interpretation by imagining situations that are not immediately present~\cite{barsalou2009simulation}. Imagining a child encountering the poster from across a hallway might foreground legibility at distance; imagining the poster outdoors could introduce contrast as a new concern. Such prospective simulations can recruit additional attributes or constraints and allow candidate interpretations to be evaluated before acting on them.

Through repeated cycles of these processes, working memory stabilizes around a situated representation rich enough to guide judgment and action. This representation remains mutable: acting on it changes the artifact or situation, which in turn provides new perceptual evidence and may cause the individual to retrieve different knowledge, reconsider constraints, or simulate new possibilities.

\subsection{What Does the Resulting Conceptual Frame Represent?}
\label{sec:concept_representation}

We call the task-specific representation produced through the above process a \frameterm{situated conceptual frame}. Such frames organize knowledge in a form that supports situated inference, planning, and action: they determine which aspects of a concept are currently relevant and how they should guide behavior in the task at hand. Note that our use of \emph{frame} differs from classical frame-based representations in AI, which typically encode relatively \textit{stable} knowledge as predefined slots, values, and defaults~\cite{minsky1974framework}. Following cognitive accounts of situated conceptualization, we instead treat a frame as a \emph{dynamically constructed, task- and situation-relative organization} of knowledge~\cite{barsalou2003situated,barsalou2009simulation,barsalou2012frames}.

Concretely, a frame organizes the aspects of a concept that are currently relevant to the task. At its core are \frameterm{attributes} and associated \frameterm{values}. For visual appeal, attributes might include color, typography, imagery, text density, or visual hierarchy, while values might include bright colors, rounded typography, cartoon imagery, reduced text, or a strong focal element. Attributes can also be hierarchical: a typography attribute may itself contain a frame with font family, weight, size, spacing, and color. Further, frame elements are are not independent. \frameterm{Relations} capture dependencies or regularities among attributes, while \frameterm{constraints} specify combinations that are permissible or required. Increasing font size, for example, may require changes to spacing or layout; playful imagery may still need to satisfy constraints on scientific accuracy. These dependencies allow a frame to represent a coherent conceptual configuration rather than a set of independent parameters.

Frames also encode \frameterm{preferred values} and \frameterm{relative importance of attributes}. A designer may seek to optimize simultaneously for playfulness, readability, and novelty while assigning greater importance to readability. Values themselves can exhibit graded structure: prior experience can make some values more typical or accessible than others, providing defaults when the situation offers little additional guidance~\cite{barsalou2012frames,barsalou1991deriving}. Thus, prior knowledge supplies distributions of possible attributes, values, and relations, while the situated frame is their momentary organization for the current task. Finally, at a broader level, frames participate in \frameterm{conceptual fields}: networks of related concepts that co-occur across situations and share systematic relations~\cite{barsalou2012frames,yeh2006situated}. The field surrounding ``visual appeal for a five-year-old,'' for instance, may activate notions such as \emph{child-friendly}, \emph{playful}, and \emph{engaging}; introducing the science-camp context may additionally activate \emph{educational}, \emph{discovery}, or \emph{outdoors}. Which broader framing is activated can reorganize which attributes, values, and relations become relevant.

We formalize this operational representation as follows.

\begin{definition}[Situated Conceptual Frame]
\label{def:frame}
For collaborator $k$, a situated conceptual frame is
\[
\mathcal{F}_k =
\left(
    A_k,\;
    \mathcal{V}_k,\;
    v_k,\;
    R_k,\;
    C_k,\;
    O_k,\;
    W_k,\;
    CF_k
\right),
\]
where:
\begin{itemize}
    \item \frameterm{$A_k$: Attributes} currently represented as relevant;
    \item \frameterm{$\mathcal{V}_k$: Values}, where
    $\mathcal{V}_k = \{\mathcal{V}_{k,a}\}_{a\in A_k}$;
    \item \frameterm{$v_k$: Instantiated or preferred values}, where
    $v_k(a) \in \mathcal{V}_{k,a}$;
    \item \frameterm{$R_k$: Relations or dependencies} among frame elements;
    \item \frameterm{$C_k$: Constraints} over permissible configurations;
    \item \frameterm{$O_k$: Optimization criteria} or preferred directions for the task;
    \item \frameterm{$W_k$: Importance or salience} of frame elements; and
    \item \frameterm{$CF_k$: Conceptual fields or framings} organizing the current interpretation.
\end{itemize}
\end{definition}

Note that this situated conceptual frame is a task-relative abstraction intended to capture the aspects of an interpretation that matter for collaborative action and repair, rather than a complete description of cognition. For humans, it is motivated by cognitive theories of situated conceptualization; for AI systems, it serves as an \frameterm{operational representation of the system's current task interpretation}, without assuming that it faithfully exposes the model's latent internal state or that AI systems possess human-like conceptual representations.

\subsection{When Are Collaborators Conceptually Misaligned?}
\label{sec:definition}

Having described how an individual constructs a situated conceptual frame, we now consider two collaborators engaged in the same task. Prior work broadly defines alignment in collaborative work as \textit{``the extent to which individuals represent things in the same way as each other''}~\cite{pickering2021understanding}. This alignment is important because it allows collaborators to predict each others action and jointly plan and act. Such alignment can occur at multiple levels, from linguistic representations to representations of the ongoing situation and conversational state~\cite{pickering2021understanding,roberts2012information}. Our focus is narrower in that \emph{conceptual alignment} concerns the \textit{compatibility} of collaborators' situated conceptual frames for guiding the joint activity. In other words, misalignment arises when collaborators construct different task-relevant interpretations of the concepts guiding their joint activity. These differences may be reflected in explicit linguistic disagreement, but they can also remain latent when collaborators use the same words while associating them with different attributes, values, relations, or priorities. We therefore define conceptual misalignment in terms of divergence between collaborators' situated conceptual frames rather than agreement or disagreement
at the level of language.

\begin{definition}[Conceptual Misalignment]
\label{def:misalignment}
Given two situated conceptual frames $\mathcal{F}_i$ and $\mathcal{F}_j$ constructed for task $T$, their conceptual divergence can be characterized as
\[
\Delta_T(\mathcal{F}_i,\mathcal{F}_j)
=
\left(
    \Delta_A,\;
    \Delta_V,\;
    \Delta_R,\;
    \Delta_C,\;
    \Delta_O,\;
    \Delta_W,\;
    \Delta_{CF}
\right)_T,
\]
where:
\begin{itemize}
    \item $\Delta_A$ captures differences in which \frameterm{attributes} are represented;
    \item $\Delta_V$ captures differences in the \frameterm{values} associated with corresponding attributes;
    \item $\Delta_R$ captures differences in \frameterm{relations or dependencies} among frame elements;
    \item $\Delta_C$ captures differences in \frameterm{constraints} over permissible configurations;
    \item $\Delta_O$ captures differences in \frameterm{optimization criteria} or preferred directions;
    \item $\Delta_W$ captures differences in relative \frameterm{importance or salience}; and
    \item $\Delta_{CF}$ captures differences in higher-level \frameterm{conceptual fields or framings}.
\end{itemize}
\end{definition}

These components describe the \emph{locus} of conceptual misalignment. Two collaborators might represent the same attribute but instantiate different values, producing $\Delta_V$. One might treat readability as relevant while the other omits it, producing $\Delta_A$. They may agree on attributes and values but disagree about how two elements depend on one another, producing $\Delta_R$, or about which configurations are permissible, producing $\Delta_C$. They may pursue different desired outcomes or assign different importance to the same considerations, producing $\Delta_O$ or $\Delta_W$, or construe the task through different broader framings, producing $\Delta_{CF}$. Tables~\ref{tab:divergence-inputs}--\ref{tab:divergence-communication} summarizes several sources of such divergence in terms of the process model developed above. A more complete discussion of these sources and their supporting literature is provided in Appendix~\ref{sec:a_variations}.


\begin{table*}[t]
\centering
\footnotesize
\setlength{\tabcolsep}{4pt}
\renewcommand{\arraystretch}{1.10}

\caption{
\textbf{Sources of divergence in conceptual inputs.}
Differences in the information available to collaborators can lead them to construct different situated conceptual frames.
}
\label{tab:divergence-inputs}

\begin{tabularx}{\textwidth}{
    p{0.19\textwidth}
    p{0.46\textwidth}
    X
}
\toprule
\rowcolor{SituationLight}
\situationterm{Source} &
\situationterm{Dimensions that can vary} &
\situationterm{Possible frame differences} \\
\midrule

\situationterm{External situation}
&
Grain size, tangibility, familiarity, and temporal, spatial, social, or hypothetical distance can alter what aspects of a situation are available or treated as relevant
~\cite{yeh2006situated,barsalou2008grounded,schank2013scripts,
logan1988toward,dane2010reconsidering,trope2010construal}.
&
Different attributes or priorities become salient ($\Delta_A,\Delta_W$), while different values, constraints, or levels of abstraction may be inferred ($\Delta_V,\Delta_C,\Delta_{CF}$).
\\[0.35em]

\situationterm{Internal task context}
&
Goals, goal abstraction, motivations, values, roles, resource constraints, and affect shape perceptual selection and what knowledge is recruited
~\cite{barsalou2018moving,hommel2001theory,barsalou1983ad,
hass2019idea,vallacher1987people,trope2003temporal,
isen1984influence,gasper2002attending,han2014emotions}.
&
Different attributes become salient ($\Delta_A,\Delta_W$), and collaborators may establish different optimization criteria, preferred values, or constraints ($\Delta_O,\Delta_V,\Delta_C$).
\\[0.35em]

\situationterm{Prior knowledge, acquisition, and expertise}
&
Different exemplars, prototypes, causal theories, episodic experiences, learning modalities, and levels of expertise provide different material for conceptual construction
~\cite{tenenbaum2011grow,posner1968genesis,rosch1975family,
medin1978context,murphy1985role,gopnik1993we,barsalou1985ideals,
medin1997categorization,barsalou1999perceptual,wauters2003mode,
villani2019varieties,juhasz2005age,barsalou2020challenges}.
&
Different candidate attributes, defaults, relations, or broader conceptual structures may be recruited ($\Delta_A,\Delta_V,\Delta_R,\Delta_{CF}$).
\\[0.35em]

\situationterm{Social and cultural experience}
&
Instruction, social learning, community norms, and cultural experience shape which distinctions are learned, which values are conceivable, and which ideals are desirable
~\cite{mervis1987child,waxman1997setters,schwartz2011practicing,
norenzayan2002cultural,naous2023having,borghi2009words,
malt1995category,kovecses2005metaphor,atran2008native,
waxman2007folkbiological,ojalehto2015perspectives,barsalou2023implications}.
&
Different defaults, acceptable values, constraints, optimization criteria, and broader framings can emerge ($\Delta_V,\Delta_C,\Delta_O,\Delta_{CF}$).
\\

\bottomrule
\end{tabularx}
\end{table*}


\begin{table*}[t]
\centering
\footnotesize
\setlength{\tabcolsep}{4pt}
\renewcommand{\arraystretch}{1.10}

\caption{
\textbf{Sources of divergence in conceptualization processes.}
Even given similar inputs, collaborators can construct different interpretations because the processes that select, retrieve, organize, and evaluate information can vary.
}
\label{tab:divergence-process}

\begin{tabularx}{\textwidth}{
    p{0.19\textwidth}
    p{0.46\textwidth}
    X
}
\toprule
\rowcolor{ProcessLight}
\processterm{Source} &
\processterm{Dimensions that can vary} &
\processterm{Possible frame differences} \\
\midrule

\processterm{Perception and attention}
&
Perceptual sensitivity, expertise, goals, attentional focus, joint attention, and cognitive load affect which information collaborators extract from the same situation
~\cite{hommel2001theory,chase1973perception,brennan1995centering,
tomasello1986joint,swallow2013attentional,xu2011gaze}.
&
Different evidence enters conceptualization, affecting represented attributes and their salience ($\Delta_A,\Delta_W$), with downstream effects on other frame components.
\\[0.35em]

\processterm{Retrieval}
&
Frequency and recency of use, contextual relevance, encoding strength, interference, and semantic-network structure affect which knowledge becomes accessible
~\cite{murphy2004big,anderson1999fan,buchanan2001characterizing,
pexman2007neural,kenett2019semantic,popov2019semantic}.
&
Different attributes, candidate values, exemplars, relations, or conceptual associations may be recruited ($\Delta_A,\Delta_V,\Delta_R,\Delta_{CF}$).
\\[0.35em]

\processterm{Reasoning}
&
Collaborators can employ different analogies, metaphors, causal models, levels of abstraction, conceptual combinations, or inferential strategies
~\cite{hummel1997distributed,thibodeau2011metaphors,
kalogerakis2010developing,ozkan2013cognitive,
chi1981categorization,Johnson2000explanatory}.
&
Different inferential paths can produce different values, relations, constraints, optimization criteria, or broader framings ($\Delta_V,\Delta_R,\Delta_C,\Delta_O,\Delta_{CF}$).
\\[0.35em]

\processterm{Simulation}
&
Collaborators may imagine different users, future situations, contexts of use, or consequences when evaluating an emerging interpretation
~\cite{barsalou2009simulation,barsalou2005situating}.
&
Different attributes, constraints, consequences, or preferred directions may become relevant ($\Delta_A,\Delta_C,\Delta_O,\Delta_W$).
\\[0.35em]

\processterm{Cognitive effort and metareasoning}
&
Working-memory capacity, depth of deliberation, confidence calibration, perceived fluency, and metacognitive control affect how extensively an interpretation is elaborated and evaluated
~\cite{hammer2019individual,unsworth2014working,
ackerman2017meta,buchel2013metacognitive,topolinski2010immediate,
kramer2023testing,reason1992cognitive,kahneman2011thinking}.
&
One collaborator may construct a more elaborated or internally constrained frame than another, producing differences across multiple components.
\\

\bottomrule
\end{tabularx}
\end{table*}


\begin{table*}[t]
\centering
\footnotesize
\setlength{\tabcolsep}{4pt}
\renewcommand{\arraystretch}{1.10}

\caption{
\textbf{Sources of divergence in concepts and frame structure.}
Properties of concepts and their learned organization influence how situated conceptual frames can be instantiated.
}
\label{tab:divergence-frame}

\begin{tabularx}{\textwidth}{
    p{0.19\textwidth}
    p{0.46\textwidth}
    X
}
\toprule
\rowcolor{FrameLight}
\frameterm{Source} &
\frameterm{Dimensions that can vary} &
\frameterm{Possible frame differences} \\
\midrule

\frameterm{Concept properties}
&
Concepts differ in concreteness, abstraction, context availability, situational systematicity, semantic diversity, associative structure, semantic richness, and hierarchical position
~\cite{barsalou2018moving,borghi2014words,borghi2017challenge,
paivio1990mental,trope2010construal,schwanenflugel1983differential,
davis2020situational,barsalou2005situating,hoffman2016meaning,
lakhzoum2021semantic,recchia2012semantic,tanaka1991object,
chi1981categorization}.
&
Some concepts constrain interpretation relatively tightly, while others permit broader differences in attributes, values, relations, and framing ($\Delta_A,\Delta_V,\Delta_R,\Delta_{CF}$).
\\[0.35em]

\frameterm{Attribute structure}
&
People vary in which attributes they associate with a concept and which properties they consider central, diagnostic, ideal, or causally explanatory
~\cite{barsalou1981instability,barsalou1993linguistic,
rosch1975family,barsalou1985ideals,murphy1985role,
Johnson2000explanatory}.
&
Collaborators may include or omit different attributes ($\Delta_A$) or assign them different relative importance ($\Delta_W$).
\\[0.35em]

\frameterm{Values and acceptable ranges}
&
Defaults, prototypes, ideals, thresholds, and acceptable ranges vary with experience, context, community, and culture
~\cite{rosch1975family,barsalou1981instability,barsalou1985ideals,
markus1991culture,medin2004native,norenzayan2002cultural}.
&
Shared attributes may be instantiated differently ($\Delta_V$), or collaborators may disagree about which values satisfy relevant constraints ($\Delta_C$).
\\[0.35em]

\frameterm{Relations and constraints}
&
People differ in causal theories, assumptions about which properties co-vary, and beliefs about which combinations are possible, permissible, or necessary
~\cite{murphy1985role,gopnik1997words,gopnik2012reconstructing,
Johnson2000explanatory}.
&
Frames may encode different dependencies ($\Delta_R$) or constraints ($\Delta_C$), causing changes to propagate differently through each frame.
\\[0.35em]

\frameterm{Priorities and optimization}
&
When desirable properties compete, collaborators can differ in which outcomes they optimize and which trade-offs they accept
~\cite{keeney1993decisions,payne1993adaptive,
barsalou1983ad,barsalou1985ideals}.
&
Collaborators may pursue different preferred directions ($\Delta_O$) or assign different importance to otherwise shared concerns ($\Delta_W$).
\\[0.35em]

\frameterm{Broader conceptual framing}
&
The same task can be situated within different conceptual fields, abstraction levels, analogical frames, or culturally learned systems of meaning
~\cite{barsalou2012frames,yeh2006situated,trope2010construal,
kovecses2005metaphor,barsalou2023implications}.
&
Different higher-level framings ($\Delta_{CF}$) can reorganize which attributes, values, relations, constraints, and priorities appear relevant.
\\

\bottomrule
\end{tabularx}
\end{table*}


\begin{table*}[t]
\centering
\footnotesize
\setlength{\tabcolsep}{4pt}
\renewcommand{\arraystretch}{1.10}

\caption{
\textbf{Sources of divergence introduced through communication.}
Even when collaborators begin with relatively compatible situated frames, communicating those frames through limited external channels can introduce additional differences.
We treat these factors as cross-cutting rather than as a fourth locus in the situated conceptualization model.
}
\label{tab:divergence-communication}

\begin{tabularx}{\textwidth}{
    p{0.19\textwidth}
    p{0.46\textwidth}
    X
}
\toprule
\rowcolor{gray!10}
\textbf{Source} &
\textbf{Dimensions that can vary} &
\textbf{Possible frame differences} \\
\midrule

\textbf{Underspecification}
&
Speakers routinely omit information they expect collaborators to recover from shared context and common ground; when this assumed overlap is weak, recipients must infer missing details
~\cite{grice1975logic,clark1996using}.
&
Different attributes, values, constraints, or priorities may be supplied during reconstruction ($\Delta_A,\Delta_V,\Delta_C,\Delta_W$).
\\[0.35em]

\textbf{Ambiguity and polysemy}
&
The same expression can support multiple interpretations, and contextual cues may be insufficient to determine which sense or conceptual frame was intended
~\cite{klein2001representation,clark1996using}.
&
Collaborators may activate different values, relations, or broader conceptual framings ($\Delta_V,\Delta_R,\Delta_{CF}$).
\\[0.35em]

\textbf{Vagueness and granularity}
&
Expressions can intentionally or unintentionally leave boundaries, thresholds, or levels of precision unspecified
~\cite{austin1975things,jucker2003interactive}.
&
Collaborators may adopt different acceptable ranges, levels of abstraction, constraints, or optimization thresholds ($\Delta_V,\Delta_C,\Delta_O,\Delta_{CF}$).
\\[0.35em]

\textbf{Compression}
&
Rich conceptual representations must be externalized through comparatively sparse linguistic, visual, or symbolic expressions, so only a subset of the underlying frame is communicated
~\cite{grice1975logic}.
&
Recipients may reconstruct omitted structure differently, producing divergence across multiple frame components.
\\[0.35em]

\textbf{Differences in common ground}
&
Collaborators can differ in what they believe is mutually known, salient, or already established in the interaction
~\cite{clark1991grounding,clark1996using,lewis1979scorekeeping}.
&
Information one collaborator treats as implicit may be absent from the other's frame, producing differences in attributes, constraints, values, or framing.
\\[0.35em]

\textbf{Channel and modality}
&
Different communicative channels provide different affordances for grounding and repair: language can express abstractions efficiently, while visual artifacts, gesture, or direct manipulation can make referents and spatial relations more explicit.
&
What can be externalized or verified differs across channels, affecting which frame components remain implicit or become jointly inspectable.
\\

\bottomrule
\end{tabularx}
\end{table*}

Importantly, divergence need not remain localized. Because frame elements are related, changing one component can produce downstream consequences for others. Revising typography from formal to playful, for instance, may alter appropriate font weight, spacing, imagery, or visual hierarchy. Conceptual repair cannot always be modeled as independently replacing one value with another; some changes require revising or propagating changes through the surrounding frame. Further, conceptual misalignment is also graded and dynamic. Collaborators may differ substantially on a single frame element or subtly across many. As they communicate, manipulate artifacts, and encounter new information, their respective frames can change, causing them to move closer together or farther apart. Misalignment is not necessarily an exceptional conversational failure, but a persistent possibility in collaborative activity~\cite{weigand1999misunderstanding}. The next question is consequently not whether collaborators' frames are identical, but whether the remaining differences are consequential enough to matter for the joint task.

\subsection{When Is Conceptual Alignment Sufficient?}
\label{sec:task-sufficient-alignment}

Successful collaboration does not require collaborators to construct identical conceptual representations. They may differ in background knowledge, prior associations, or other aspects of their situated frames while still making compatible judgments and coordinating their actions. What matters is whether the differences that remain are consequential for the joint activity at hand. A relatively small disagreement about a critical safety requirement, accessibility constraint, or design priority may substantially alter what collaborators decide or produce, whereas much larger differences in peripheral associations may have no effect on their ability to proceed together. Conceptual alignment is thus \emph{task sufficient} alignment.

\section{Alignment Games: How Can Collaborators Repair Conceptual Misalignment?}
\label{sec:alignment-games}

Because situated frames evolve as collaborators act and receive feedback, alignment is likewise an ongoing \textit{interactive} process.  Our account builds on a long tradition of treating communication as collaborative action. Speech-act theory characterizes utterances in terms of the effects speakers intend to produce~\cite{austin1975things,searle1969speech}; grounding and common-ground theories describe how collaborators establish and repair shared understanding~\cite{clark1991grounding,clark1996using}; and interactive alignment accounts explain how interlocutors converge across linguistic and situational representations~\cite{garrod2004conversation,pickering2021understanding}. Related work shows that collaborators form local conceptual pacts~\cite{brennan1996conceptual,ibarra2016flexibility} and coordinate through artifacts that can accommodate multiple interpretations~\cite{star1989structure,bowker2000sorting,lee2005between}. Most directly, \emph{dialog games} formalize recurring patterns of communicative initiative, response, and repair~\cite{mann1988dialogue,hulstijn2000dialogue}.

Alignment Games extend these traditions by shifting the object of coordination from whether an utterance has been understood to the \emph{task-specific conceptualization guiding action}. Collaborators may share a referent and understand the same words while still disagreeing about which attributes matter, what values they should take, how they relate, or which considerations should be prioritized. Alignment Games describe the interactional work through which these conceptual differences become inspectable and repairable.

\subsection{What Is an Alignment Game?}
\label{sec:what-is-game}

\begin{definition}[Alignment Game]
An \emph{Alignment Game} is a structured, multi-turn interaction through which collaborators diagnose, negotiate, repair, and validate task-relevant differences between their situated conceptual frames.
\end{definition}

An Alignment Game begins when collaborators encounter evidence that
their current conceptualizations may not support coordinated action in the
situation at hand. This need not mean that they are misaligned with respect to the entire task, a difference may become consequential only for a particular decision or subtask. Evidence of misalignment can be \emph{direct}, as when a collaborator says, ``I do not think we mean the same thing by \textit{playful}.'' It can also be \emph{indirect}, as when a collaborator evaluates an artifact as ``too childish,'' produces an unexpected result, or takes an action that conflicts with the other's expectations. Such evidence makes some task-relevant difference in their situated frames available for inspection. Collaborators respond through one or more \emph{alignment moves}: actions intended to reveal, modify, or test some part of a collaborator's task-relevant conceptualization. An alignment move may be linguistic---for example, asking what ``playful'' means, proposing a more specific attribute, or stating a preference---but it need not be a speech act. Collaborators can also point to an example, modify an artifact, compare alternatives, or enact a possible interpretation. What distinguishes an alignment move is its function rather than its communicative form; it is performed to affect the
conceptual representation guiding subsequent joint action.

\begin{figure*}[t]
    \centering
    \includegraphics[width=\linewidth]{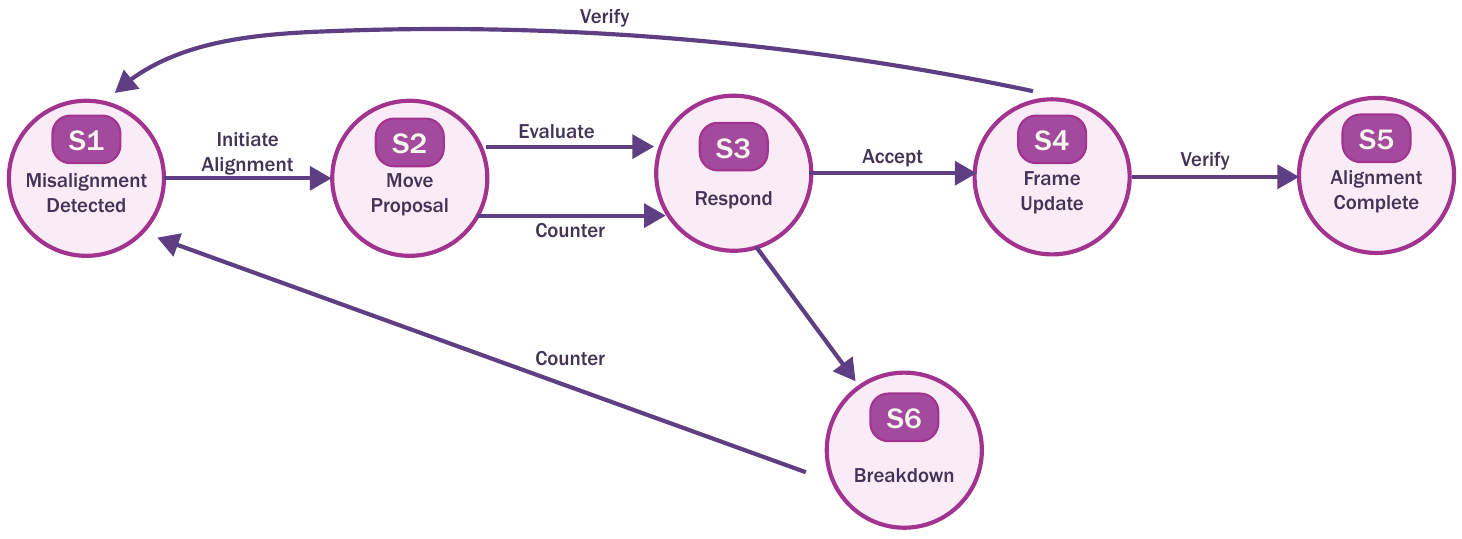}
    \caption{
    \textbf{The Alignment Game cycle.}
    Evidence from an utterance, action, or artifact can reveal a possible conceptual difference. A collaborator responds with an alignment move intended to diagnose, repair, or validate that difference. The partner's uptake---acceptance, rejection, counterproposal, or partial enactment---provides new evidence and may update one or both situated frames. The cycle continues until the remaining difference is sufficiently small for the joint task or the repair is abandoned.
    }
    \Description{
    State diagram of six circular nodes. S1 Misalignment Detected leads via "Initiate Alignment" to S2 Move Proposal, then via "Evaluate" or "Counter" to S3 Respond. From S3, "Accept" leads to S4 Frame Update, which "Verify" leads to S5 Alignment Complete or loops back to S1. S3 can also fall to S6 Breakdown, which returns to S1 via "Counter."
    }
    \label{fig:alignment-game}
\end{figure*}

As shown in Figure~\ref{fig:alignment-game}, a move need not immediately produce agreement. Its addressee may accept it, reject it, counter it, reinterpret it, or enact it only partially. The resulting response provides new evidence from which the collaborators update their respective frames and decide whether additional repair is necessary. The game proceeds iteratively through a recurring pattern of \emph{evidence of divergence $\rightarrow$ diagnosis or repair $\rightarrow$ uptake $\rightarrow$ frame update $\rightarrow$ validation}. It ends when the remaining difference is task-sufficient, or when the collaborators abandon or defer the repair.

An Alignment Game is successful when collaborators reach the task-sufficient state defined in Section~\ref{sec:task-sufficient-alignment}; peripheral differences may remain. Moreover, convergence need not be one-sided. One collaborator may revise toward the other's interpretation, both may change, or interaction with the evolving artifact may lead both toward a new interpretation that neither initially held.

\subsection{What Is an Alignment Move?}
\label{sec:alignment-move}

An \emph{alignment move} is the basic functional unit of an Alignment Game. Unlike a dialog act~\cite{mann1988dialogue}, which characterizes the communicative function of an utterance (e.g., asking, informing, confirming), an alignment move is defined by its intended effect on a collaborator's \emph{situated conceptual frame}. Specific moves can also have preconditions. An example is useful only if the partner can interpret the exemplar; a direct value repair presupposes that the relevant attribute is already shared; and an analogy requires sufficient knowledge of the source domain. Whether a move succeeds depends not only on its theoretical fit to the misalignment but also on whether the partner can recognize and take up the intervention.

We characterize an alignment move along five dimensions:

\begin{itemize}
    \item \textbf{Function}: whether the move primarily \emph{diagnoses}, \emph{repairs}, \emph{negotiates}, or \emph{validates} a conceptual difference;
    \item \textbf{Locus}: whether it intervenes on the \situationterm{situation}, the \processterm{construction process}, or the \frameterm{conceptual frame};
    \item \textbf{Target}: the particular information, process, or frame component the move addresses;
    \item \textbf{Scope}: whether the intervention is local to a single element, spans several related elements, or reorganizes the interpretation more broadly; and
    \item \textbf{Expected effect}: the difference the move is intended to reveal or reduce, such as $\Delta_A$, $\Delta_V$, $\Delta_R$, $\Delta_C$, $\Delta_O$, $\Delta_W$, or $\Delta_{CF}$.
\end{itemize}

We organize alignment moves by the \emph{locus at which they intervene} in the situated conceptualization model. This organization connects the interaction framework directly to the cognitive account in Section~\ref{sec:situated-conceptualization}: a repair can alter the information entering conceptualization, the processes used to construct an interpretation, or the resulting conceptual frame itself. 

\subsubsection{Situation-Level Moves: Changing the Basis for Conceptualization}
\label{sec:situation-moves}

\situationterm{Situation-level moves} intervene on the information from which a conceptualization is constructed. They can establish a shared referent, add missing evidence, narrow attention to relevant parts of the situation, clarify goals or constraints, or introduce examples that provide additional grounding. Rather than directly specifying how the partner's frame should change, these moves alter the evidence or task context from which the partner constructs it. Situation-level repair is particularly useful when collaborators may be reasoning from different artifacts, instructions, contexts, or assumptions. For example, ``Ignore the footer; I mean the title area'' changes which part of the artifact is treated as relevant, while ``This will be viewed from across a school hallway'' introduces contextual information that can subsequently alter attributes such as text size and contrast.

\begin{table*}[t]
\centering
\footnotesize
\setlength{\tabcolsep}{4pt}
\renewcommand{\arraystretch}{1.10}

\caption{
\textbf{Situation-level alignment moves.}
These moves intervene on situational and task information that serves as input to situated conceptualization.
}
\label{tab:situation-moves}

\begin{tabularx}{\textwidth}{
    p{0.20\textwidth}
    p{0.16\textwidth}
    p{0.34\textwidth}
    X
}
\toprule
\rowcolor{SituationLight}
\situationterm{Move} &
\situationterm{Function} &
\situationterm{Target / Intended effect} &
\situationterm{Example} \\
\midrule

\situationterm{Ground$(e)$}
&
Repair, validate
&
\textbf{Situational elements:} establishes a shared referent or directs collaborators to the same observable element.
&
\textit{``This icon is the mascot.''}
\\[0.3em]

\situationterm{Augment$(e)$}
&
Repair
&
\textbf{Situation / context:} introduces missing information or resources that should enter conceptualization.
&
\textit{``Here is the brand palette.''}
\\[0.3em]

\situationterm{Filter$(e,\mathrm{scope})$}
&
Repair
&
\textbf{Situational elements:} narrows the relevant evidence or excludes information that should not influence the interpretation.
&
\textit{``Ignore the footer; focus on the title area.''}
\\[0.3em]

\situationterm{SpecifyContext$(g,r,c)$}
&
Diagnose, repair, validate
&
\textbf{Goals, roles, constraints:} makes relevant task context explicit so collaborators reason from compatible conditions.
&
\textit{``Brand compliance is required, and this is for five-year-olds.''}
\\[0.3em]

\situationterm{ProvideSchema$(f)$}
&
Repair
&
\textbf{Situation--frame coupling:} provides an organizing schema through which available information can be interpreted.
&
\textit{``Read this as a Z-layout: title at the top, image in the center, caption below.''}
\\[0.3em]

\situationterm{Simulate$(scenario)$}
&
Diagnose, repair
&
\textbf{Prospective situation:} introduces an imagined condition or context that can reveal additional constraints, goals, or consequences.
&
\textit{``Picture this being viewed on a billboard outdoors at noon.''}
\\[0.3em]

\situationterm{Flag$(e,status)$}
&
Repair
&
\textbf{Element status:} marks an observable element as discrepant, satisfactory, or otherwise relevant to the mismatch.
&
\textit{``This color is too dull.''}
\\[0.3em]

\situationterm{Example$(+)$}
&
Diagnose, repair, validate
&
\textbf{Exemplar / values:} provides a positive anchor from which intended attributes or values can be inferred.
&
\textit{``Like these children's museum posters---bright, playful, but still clear.''}
\\[0.3em]

\situationterm{CounterExample$(-)$}
&
Diagnose, validate
&
\textbf{Negative exemplar / values:} rules out an interpretation or region of the conceptual space.
&
\textit{``Not like a Halloween poster; avoid muted grays.''}
\\[0.3em]

\situationterm{SpecifyOutput$(spec)$}
&
Validate
&
\textbf{Output criteria:} makes externally verifiable requirements for the resulting artifact explicit.
&
\textit{``Let's target at least 4.5:1 contrast for body text.''}
\\

\bottomrule
\end{tabularx}
\end{table*}

\subsubsection{Process-Level Moves: Changing How the Frame Is Constructed}
\label{sec:process-moves}

\processterm{Process-level moves} intervene on the cognitive operations through which available information is transformed into a situated frame. They can redirect attention, cue retrieval, encourage decomposition or abstraction, invoke analogy, or ask the collaborator to simulate a possible situation. These moves are useful when the desired change is difficult to express as a direct edit to a frame component or when collaborators need to explore alternative ways of construing the task. For example, ``Imagine a five-year-old seeing this from across the hallway'' does not directly prescribe a font size or contrast value. Instead, it recruits simulation, which may cause those attributes and constraints to become salient. Similarly, ``Think of this more like a children's museum than a birthday party'' can invoke analogical reasoning that reorganizes several parts of the frame at once.

\begin{table*}[t]
\centering
\footnotesize
\setlength{\tabcolsep}{4pt}
\renewcommand{\arraystretch}{1.10}

\caption{
\textbf{Process-level alignment moves.}
These moves intervene on reasoning processes used to construct, elaborate, reorganize, or evaluate situated conceptual frames.
}
\label{tab:process-moves}

\begin{tabularx}{\textwidth}{
    p{0.20\textwidth}
    p{0.16\textwidth}
    p{0.34\textwidth}
    X
}
\toprule
\rowcolor{ProcessLight}
\processterm{Move} &
\processterm{Function} &
\processterm{Target / Intended effect} &
\processterm{Example} \\
\midrule

\processterm{Decompose$(a)$}
&
Repair
&
\textbf{Attributes ($A$):} splits an overloaded or coarse attribute into more specific sub-attributes that can be reasoned about separately.
&
\textit{``Decompose typography into family, weight, size, and spacing.''}
\\[0.3em]

\processterm{Abstract$(level\,\uparrow/\downarrow)$}
&
Repair
&
\textbf{Attributes / values:} changes the level of abstraction or granularity at which the concept is being reasoned about.
&
\textit{``Not navy versus teal---think warm versus cool palette.''}
\\[0.3em]

\processterm{Restructure$(h)$}
&
Repair
&
\textbf{Frame / hierarchy:} reorganizes the structure or ordering of frame elements.
&
\textit{``Shift to a type-first layout; text leads and the image supports.''}
\\[0.3em]

\processterm{AnalogicalMap$(src)$}
&
Probe, repair
&
\textbf{Relations / values:} transfers relational structure from a familiar source domain to reorganize the current interpretation.
&
\textit{``Lay it out like a metro map: line colors are categories and stops are sections.''}
\\[0.3em]

\processterm{BlendConcepts$(c_1,c_2)$}
&
Repair, create
&
\textbf{Values / conceptual fields:} synthesizes selected structures or values from two conceptual sources.
&
\textit{``Blend magazine minimalism with children's-book playfulness.''}
\\[0.3em]

\processterm{Elaborate$(detail)$}
&
Repair
&
\textbf{Attributes / values:} expands an underspecified concept by generating additional detail or distinctions.
&
\textit{``Define what you mean by `playful'.''}
\\[0.3em]

\processterm{Counterfactual$(s')$}
&
Probe
&
\textbf{Constraints / optimization:} stress-tests an interpretation under an altered hypothetical condition.
&
\textit{``What if it had to work in black and white only?''}
\\[0.3em]

\processterm{TeachHeuristic$(h)$}
&
Teach, repair
&
\textbf{Optimization / constraints:} provides a reusable rule or procedure for constructing or evaluating a frame.
&
\textit{``When clarity fails, check scale, then weight, then spacing, then color.''}
\\[0.3em]

\processterm{WorkedExample$(ex\rightarrow proc)$}
&
Teach, probe
&
\textbf{Frame-construction process:} demonstrates a construction procedure step by step so that it can be reproduced or inspected.
&
\textit{``Start with one focal element, add subheads, then details, and verify with the squint test.''}
\\[0.3em]

\processterm{ThinkAloudDemo$(proc)$}
&
Teach
&
\textbf{Reasoning / evaluation process:} externalizes tacit criteria or reasoning steps that would otherwise remain implicit.
&
\textit{``I check alignment by tracing invisible columns before adjusting color.''}
\\

\bottomrule
\end{tabularx}
\end{table*}

\subsubsection{Frame-Level Moves: Directly Revising the Conceptual Frame}
\label{sec:frame-moves}

\frameterm{Frame-level moves} intervene directly on the situated conceptual frame. Once collaborators have localized a difference, they can add or remove attributes, propose values, revise dependencies, establish constraints, change priorities, or adopt a different broader framing. These moves provide the most direct correspondence between interaction and the components introduced in Definition~\ref{def:frame}. Importantly, a frame-level move is not necessarily a simple parameter update. Because frame elements can depend on one another, changing one component may require coordinated downstream changes. Setting typography from formal to playful, for example, may change appropriate font weight, spacing, imagery, or hierarchy. Frame-level moves can be local or propagate through a larger portion of the representation.

\begin{table*}[t]
\centering
\footnotesize
\setlength{\tabcolsep}{4pt}
\renewcommand{\arraystretch}{1.10}

\caption{
\textbf{Frame-level alignment moves.}
These moves intervene directly on components of the situated conceptual frame.
}
\label{tab:frame-moves}

\begin{tabularx}{\textwidth}{
    p{0.20\textwidth}
    p{0.16\textwidth}
    p{0.34\textwidth}
    X
}
\toprule
\rowcolor{FrameLight}
\frameterm{Move} &
\frameterm{Function} &
\frameterm{Target / Intended effect} &
\frameterm{Example} \\
\midrule

\frameterm{ScopeAttribute$(a)$}
&
Repair, expand
&
\textbf{Attributes ($A$):} introduces an attribute that should be represented as relevant, reducing an attribute-level mismatch.
&
\textit{``Let's add texture as an attribute.''}
\\[0.3em]

\frameterm{PruneAttribute$(a)$}
&
Repair
&
\textbf{Attributes ($A$):} removes an attribute that is irrelevant, redundant, or unnecessarily represented.
&
\textit{``We don't need to track border width separately.''}
\\[0.3em]

\frameterm{Group$([a])$}
&
Repair
&
\textbf{Attributes ($A$):} coordinates several attributes as a meaningful conceptual bundle.
&
\textit{``Treat color, type, and icon style together as one playfulness bundle.''}
\\[0.3em]

\frameterm{CalibrateMetric$(a,m)$}
&
Repair, validate
&
\textbf{Values / measurement:} aligns how variation along an attribute is represented or measured.
&
\textit{``Font size here means points, not pixels.''}
\\[0.3em]

\frameterm{ProposeValue$(a,v)$}
&
Elicit, repair
&
\textbf{Values ($V$):} proposes a candidate value for a shared attribute.
&
\textit{``The header color should be deep blue.''}
\\[0.3em]

\frameterm{Relate$(a_i,a_j)$}
&
Repair
&
\textbf{Relations ($R$):} establishes or revises a dependency between frame elements.
&
\textit{``Line height should increase when font size increases.''}
\\[0.3em]

\frameterm{ProbeValue$(a,v_1,v_2)$}
&
Probe, repair
&
\textbf{Values ($V$):} tests a distinction between candidate values to localize the intended setting.
&
\textit{``Which works better: this font weight or the lighter one?''}
\\[0.3em]

\frameterm{InduceVal$(a,ex\rightarrow r)$}
&
Repair, validate
&
\textbf{Values ($V$):} uses exemplars to infer an intended value or acceptable value range.
&
\textit{``Target colors typical of vintage science prints.''}
\\[0.3em]

\frameterm{SetConstraint$(c)$}
&
Repair, validate
&
\textbf{Constraints ($C$):} specifies a requirement governing permissible frame configurations.
&
\textit{``Contrast must be at least 4.5:1 for body text.''}
\\[0.3em]

\frameterm{SetOptimization$(o)$}
&
Diagnose, repair
&
\textbf{Optimization criteria ($O$):} specifies the preferred direction in which the task should be optimized.
&
\textit{``Prioritize readability over novelty.''}
\\[0.3em]

\frameterm{MapField$(cf)$}
&
Repair, probe
&
\textbf{Conceptual field ($CF$):} anchors the interpretation within a broader conceptual region or family.
&
\textit{``Think of this as playful minimalism.''}
\\[0.3em]

\frameterm{Reframe$(f')$}
&
Repair
&
\textbf{Frame / conceptual framing:} reorganizes the interpretation around a different higher-level frame.
&
\textit{``Treat this as a campaign poster, not a flyer.''}
\\[0.3em]

\frameterm{SetWeight$(a,w)$}
&
Repair, validate
&
\textbf{Importance ($W$):} changes the relative importance assigned to a represented concern.
&
\textit{``Accessibility should count more than aesthetics.''}
\\[0.3em]

\frameterm{Distinctor$(a)$}
&
Repair
&
\textbf{Conceptual field / frame:} uses a distinguishing attribute to constrain the conceptual perspective through which the concept is interpreted.
&
\textit{``Interpret `playful' specifically through the lens of visual simplicity.''}
\\

\bottomrule
\end{tabularx}
\end{table*}

Note that the three families are complementary. A situation-level move can cause a frame-level update, and a process-level move can reorganize several frame elements simultaneously. We classify a move according to \emph{where the collaborator intervenes}, not according to every downstream consequence that intervention may produce. Further, individual alignment moves become \emph{games} when they are taken up and combined across turns. We do not assume a single canonical sequence. Rather, different patterns become useful depending on what collaborators currently know about the misalignment and how the partner responds.

\subsection{What Factors Inform Move Selection and Repair Effort?}
\label{sec:move-selection}

The utility of any move at a given moment depends not only on the underlying divergence, but also on what collaborators currently know about it and how much coordinated change is required to resolve it. We characterize this variation along three complementary dimensions: \emph{diagnostic specificity}, \emph{task significance}, and \emph{repair effort}.

\subsubsection{Diagnostic specificity}
The first dimension captures how precisely a collaborator has localized the mismatch. At one extreme, they may only know that ``something feels off.'' At the other, they may identify a particular frame element, such as the value of \emph{typography} or the relative priority of \emph{readability} and \emph{novelty}. Diagnostic specificity depends on what evidence about the partner's interpretation is available. When only an artifact or response is visible, collaborators may have to infer the underlying source of the mismatch from its effects. Interfaces that expose portions of an operational frame can make this diagnosis more direct by revealing which attributes, values, relations, or
priorities are currently shaping the system's interpretation. When specificity is low, a useful alignment move may seek additional evidence before attempting repair.

\subsubsection{Task Significance}
The second dimension captures whether a conceptual difference is consequential for the current joint activity. The extent of representational difference and its task significance need not coincide. Collaborators may differ substantially in background associations or peripheral aspects of their situated frames while still making compatible judgments and coordinating their actions. Conversely, a relatively small disagreement about a critical constraint or priority may substantially change what they produce, evaluate, or do next. Task significance can also change during an Alignment Game. A difference that is irrelevant while collaborators are establishing the broad direction of a design may become consequential when they must select a particular representation, evaluate an artifact, or choose how to proceed.

\subsubsection{Repair Effort}
The third dimension captures how much interaction is required to move from the current misalignment to a task-sufficient state. Some differences can be repaired locally. If both collaborators already represent \emph{typography} as relevant but disagree only about its preferred value, for example, a single contrast or value-setting move may be sufficient. Other differences require broader repair. Reframing a poster from a ``birthday invitation'' to ``children's educational material,'' for instance, may alter several elements of the situated frame at once, including which attributes are relevant, what values they take, how they relate, and which considerations are prioritized.

Repair effort depends on both the number and difficulty of moves required. Moves can differ in how difficult they are to formulate, interpret, coordinate, and incorporate. Repair can also introduce \emph{collateral change}. A move may correct the targeted mismatch while unnecessarily altering parts of the frame that were already satisfactory. Effective repair seeks a task-sufficient state with as little unnecessary change as possible.

\subsubsection{Move selection under uncertainty}
Together, the above dimensions shape which move is useful at a particular point in an Alignment Game. A move may be valuable because it helps localize the mismatch, reduces a difference that matters for the task, or reaches a task-sufficient state with relatively little additional effort. 

These considerations are not intended as quantities that collaborators
explicitly optimize. Rather, they characterize the tradeoffs that arise during conceptual repair. When diagnostic specificity is low, moves that produce better evidence about a collaborator's interpretation may be more useful than moves that immediately attempt to change it. Once the mismatch has been localized, repair can focus on the differences that matter for the current task. Among otherwise adequate repairs, collaborators may favor moves that require less coordination and introduce fewer unnecessary changes to parts of the frame that are already compatible.

The usefulness of a move depends on both the current state of misalignment and the evidence available about it. Making an operational frame observable may improve specificity by exposing where the system's interpretation differs, while move scaffolds may help collaborators translate that diagnosis into an appropriate intervention. In our framework, the three move families characterize \emph{where and how} collaborators can intervene, whereas diagnostic specificity, task significance, and repair effort characterize the conditions that shape \emph{which intervention is useful at a particular moment}.
\section{Alignment Games in Existing Human--AI Interfaces}
\label{sec:examples}

\begin{figure*}
    \centering
    \includegraphics[width=\linewidth]{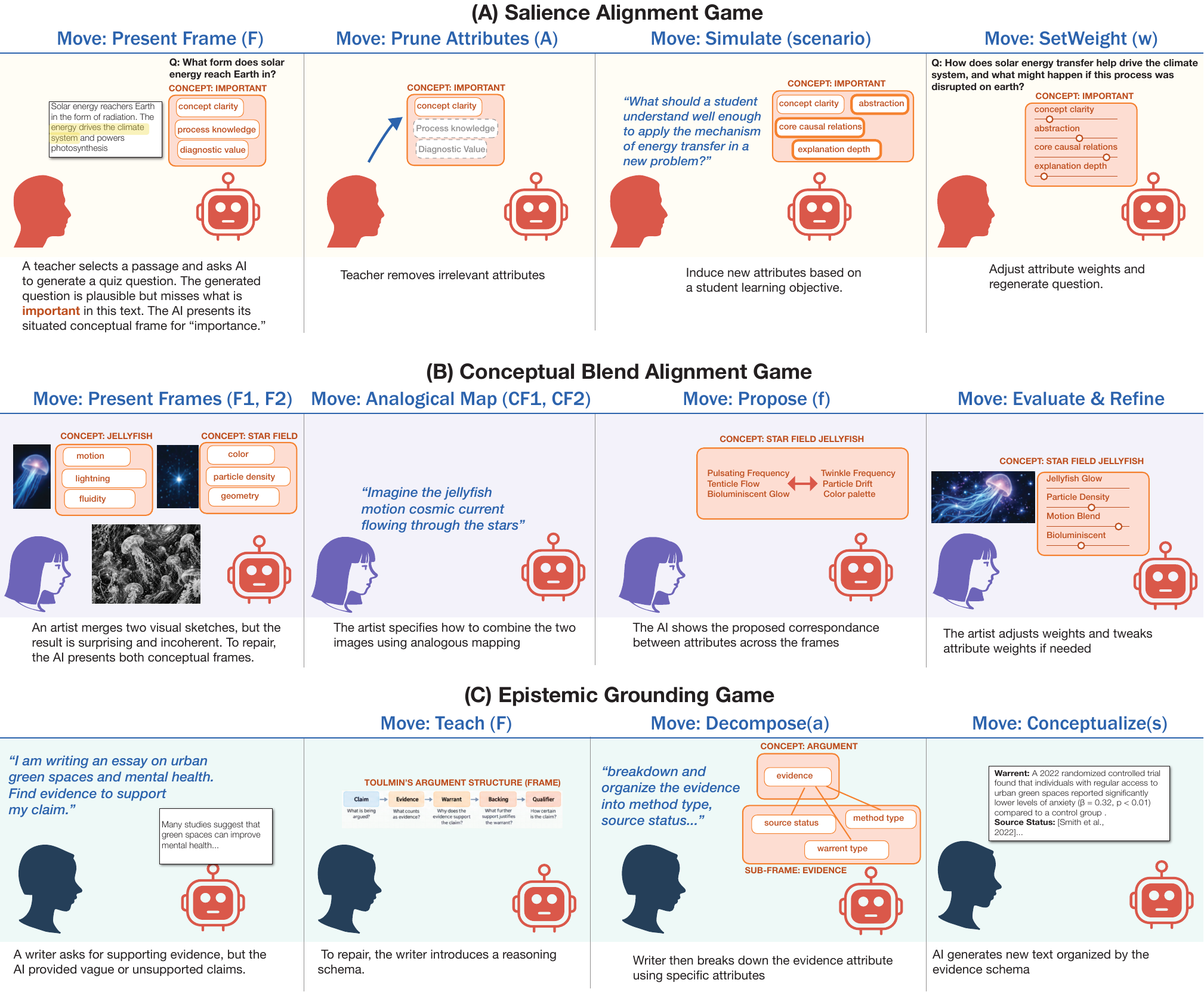}
    \caption{
    Example Alignment Games. 
    }
    \Description{Three rows of comic-style panels, each showing a person and a robot exchanging alignment moves. Row A, a teacher and AI refine what "important" means in a solar-energy quiz question. Row B, an artist and AI blend a jellyfish sketch with a star field. Row C, a writer and AI structure evidence for an essay using a Toulmin argument frame.}
    \label{fig:examples}
\end{figure*}

To illustrate the applicability of Alignment Games, we revisit breakdowns reported in three HCI systems spanning educational content generation, creative coding, and argumentative writing (Figure~\ref{fig:examples}). We interpret each breakdown through our framework and sketch a plausible sequence of alignment moves that could support its diagnosis and repair. Note that these examples are \emph{analytic reinterpretations}, not evaluations of the original systems or claims about the actual internal representations of their users or models. Instead, they ask where a consequential conceptual difference might lie, which moves could address it, and what interface affordances would make those moves easier to perform.

\subsection{Salience Alignment Game}

In an AI-assisted quiz-generation system~\cite{lu2023readingquizmaker}, teachers select textbook content from which an AI generates questions. The evaluation reports cases in which question quality suffered because it was ``hard for the AI to identify what to focus on.'' One plausible interpretation is a conceptual mismatch around what counts as \emph{important} in the selected text. A system might emphasize surface cues such as entity frequency or discourse position, whereas an instructor might prioritize causal mechanisms, central claims, or ideas that students should be able to apply. The resulting difference may therefore involve both the attributes treated as relevant ($\Delta_A$) and their relative priority ($\Delta_W$).

A \emph{Salience Alignment Game} could proceed through the following moves:

\begin{itemize}
    \item \textbf{Present(F):} The system exposes its operational frame for \emph{importance}, including candidate attributes such as \emph{entity frequency}, \emph{discourse position}, \emph{lexical novelty}, and \emph{rhetorical role}. Making these
    attributes visible helps the instructor localize how the system is currently determining what deserves attention.

    \item \textbf{Prune(A):} The instructor removes attributes that should not drive question generation, such as \emph{entity frequency} or \emph{lexical novelty}.

    \item \textbf{Simulate(scenario):} The instructor asks the system to reconsider importance relative to a learning outcome---for example, ``What should a student understand well enough to apply the mechanism of energy transfer in a new problem?'' The simulation may surface additional attributes such as \emph{core causal relations}, \emph{explanation depth},
    or \emph{transfer relevance}.

    \item \textbf{SetWeight(w):} The instructor adjusts the relative importance of these attributes, for example increasing the priority of causal relations while reducing that of rhetorical position. The revised operational frame then conditions subsequent question generation.
\end{itemize}

Here, the moves shift interaction from repeatedly regenerating questions to directly negotiating what \emph{importance} means for the instructional task. \textbf{Present} supports diagnosis, while \textbf{Prune}, \textbf{Simulate}, and \textbf{SetWeight} provide increasingly targeted means of repairing the operative conceptualization.

\subsection{Conceptual Blend Alignment Game}

Spellburst supports exploratory creative coding through natural-language prompts and allows users to semantically merge elements from different sketches~\cite{angert2023spellburst}. Its evaluation reports that some merge operations produced results participants found surprising or incoherent. One possible interpretation is that the user and system differed in what
properties of the source sketches should participate in the concept of \emph{merging}. Such a mismatch could involve the conceptual fields activated by each sketch ($\Delta_{CF}$), which attributes are carried into the blend ($\Delta_A$), and their relative importance ($\Delta_W$).

A \emph{Conceptual Blend Alignment Game} could proceed as follows:

\begin{itemize}
    \item \textbf{Present(F1, F2):} When a merge is initiated, the system presents operational frames for the two source sketches. For example, one frame might expose \emph{geometry}, \emph{motion}, and \emph{lighting} for a jellyfish sketch, while the other exposes \emph{particle density}, \emph{motion}, and \emph{color palette} for a star-field sketch. This makes visible which properties could potentially participate in the merge.

    \item \textbf{AnalogicalMap(CF1, CF2):} The user specifies a desired relation between the sources---for example, ``imagine the jellyfish motion as a cosmic current flowing through the stars.'' Alternatively, the user could explicitly select properties to blend or prune properties that should not transfer.

    \item \textbf{Propose(f):} The system proposes candidate correspondences, such as mapping \emph{pulsation frequency} to \emph{twinkle frequency} or \emph{tentacle flow} to \emph{particle drift}, and previews the
    resulting blend.

    \item \textbf{Evaluate(v):} The user inspects the proposal and resulting sketch. If consequential differences remain, they can continue the game with moves such as \textbf{SetWeight} to reduce the influence of an unwanted property or \textbf{Prune} to remove it entirely.
\end{itemize}

This example illustrates repair as a trajectory rather than a single correction. \textbf{Present} helps localize what the system is carrying across the merge; \textbf{AnalogicalMap} makes the intended relationship explicit; and \textbf{Propose} and \textbf{Evaluate} allow that interpretation to be tested before further repair. Such support can also reduce collateral change
by making alternative mappings inspectable before they are committed.

\subsection{Epistemic Grounding Game}

VIZAR is an AI-assisted writing system that provides ``argumentation sparks'' for expanding or strengthening an argumentative
essay~\cite{zhang2023visar}. Participants reported that some suggestions were too vague or insufficiently supported; for example, a request for supporting evidence could produce a statement such as ``there is scientific research to prove this'' without specifying the source or how it supports the claim. One possible interpretation is a mismatch in what counts as adequate
\emph{supporting evidence} for the current writing task. Here, the divergence may involve the constraints governing acceptable evidence ($\Delta_C$), the values instantiated for evidence type or source ($\Delta_V$), and the relative priority assigned to criteria such as rigor and persuasiveness ($\Delta_W$).

An \emph{Epistemic Grounding Game} could involve:

\begin{itemize}
    \item \textbf{Teach(c):} The user provides a reasoning structure that should organize the system's interpretation of evidence---for example, Toulmin's sequence of Claim $\Rightarrow$ Evidence $\Rightarrow$ Warrant $\Rightarrow$ Backing $\Rightarrow$ Qualifier. Rather than merely asking for ``better evidence,'' this move makes explicit the epistemic roles that supporting information should satisfy.

    \item \textbf{Decompose(a):} The user decomposes \emph{evidence} into more specific attributes, such as \emph{source status},
    \emph{method type}, \emph{effect size}, \emph{recency}, and \emph{warrant type}. These attributes provide a more explicit
    representation of the criteria through which candidate evidence should be interpreted and evaluated.

    \item \textbf{Conceptualize(s):} The system reconstructs its operational interpretation of adequate supporting evidence using this structure and generates an argumentation spark consistent with the revised frame.
\end{itemize}

Unlike the previous examples, the repair here is not primarily a matter of selecting a different value. The user changes the conceptual structure and constraints through which the system interprets \emph{evidence}. 
\section{Discussion}
\label{sec:discussion}

Alignment Games shifts the unit of analysis in human--AI interaction from whether an output satisfies a user's goal to how collaborators arrive at a task-sufficient interpretation of what should be produced. Together with the examples in Section~\ref{sec:examples}, the framework suggests implications for how generative interfaces can make conceptual alignment more explicit and designable.

\subsection{Design Principles for Alignment Games}
Building on the example vignettes in Section~\ref{sec:examples}, alignment Games makes conceptual repair an explicit object of interface design. Our framework suggests four design principles for helping collaborators diagnose consequential differences, perform targeted repairs, and reach task-sufficient alignment with less interactional effort.

\textbf{P1: Make the Operative Conceptualization Observable and Operable.} A basic challenge in conceptual repair is that users often see the \emph{product} of an AI's interpretation without seeing the interpretation
that produced it. Most generative interfaces organize interaction around a prompt--output loop, i.e., users provide an instruction, inspect an artifact, and revise the prompt or artifact until the result is acceptable
~\cite{gao2024taxonomy,luera2024survey}. When a result is unexpected, users must work backward from the artifact to infer whether the system selected the wrong attribute, instantiated an unexpected value, omitted a constraint, prioritized the wrong consideration, or framed the task differently.

Alignment Games suggests moving interaction upstream from \emph{prompt-to-output} toward \emph{prompt-to-frame}: treating the system's operative interpretation as an interaction object that can be inspected and
revised. Such a representation might expose relevant attributes and values, relations, constraints, priorities, exemplars, or broader conceptual framings. Importantly, the operational frame is not intended as a faithful readout of a model's internal cognition. Rather, it functions as a task-relative \emph{boundary object} through which collaborators can inspect, question, and revise the interpretation guiding generation
~\cite{star1989structure}.

\begin{figure*}
    \centering
    \includegraphics[width=\linewidth]{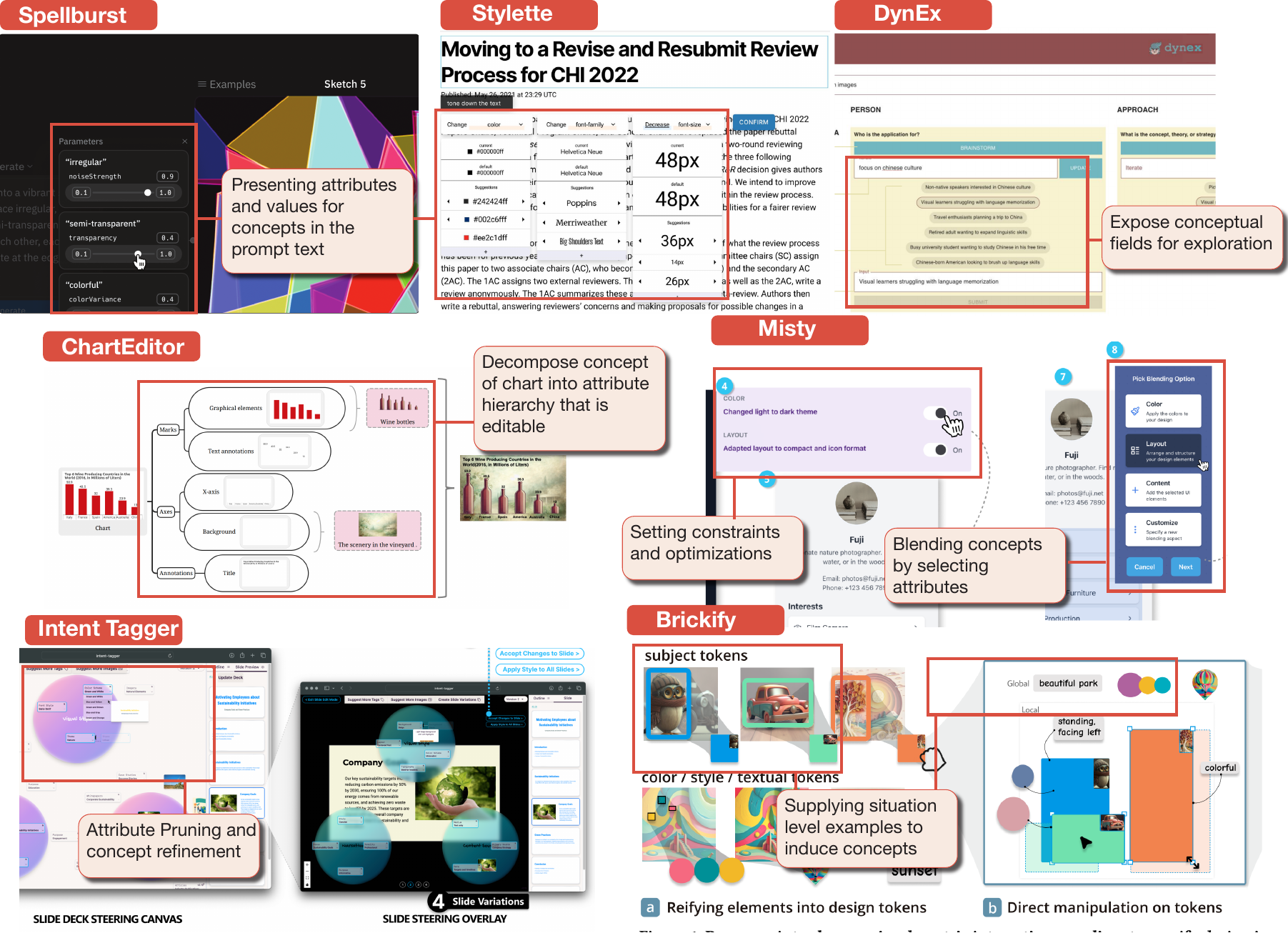}
    \caption{
        Examples of existing systems that expose conceptual frame elements and support conceptual alignment interactions. 
    }
    \Description{Collage of six annotated screenshots from prior systems, each named in an orange tab: Spellburst, Stylette, DynEx, ChartEditor, Misty, Intent Tagger, and Brickify. Callouts highlight how each exposes frame elements, such as presenting attributes and values in prompts, decomposing a chart into an editable attribute hierarchy, setting constraints, blending concepts by selecting attributes, and supplying examples to induce concepts.}
    \label{fig:systems}
\end{figure*}

Existing systems already instantiate parts of this principle (Figure~\ref{fig:systems}). \textit{Stylette}~\cite{kim2022stylette} and
\textit{Spellburst}~\cite{angert2023spellburst} expose attributes or values that users can inspect or manipulate; \textit{ChartEditor}~\cite{yan2025charteditor} makes hierarchical conceptual structure editable; and systems such as \textit{DynEx}~\cite{ma2025dynex} and \textit{Misty}~\cite{lu2025misty}, Intent Tagger~\cite{gmeiner2025intent}, and Brickify~\cite{shi2025brickify} expose broader conceptual fields for exploration. Viewed through Alignment Games, these mechanisms make different parts of an operative conceptualization available for diagnosis or
intervention. Because conceptual frames may themselves be hierarchical and interconnected, interfaces may also need to support movement between levels of abstraction, from adjusting a local value to adding an attribute, modifying a relation or constraint, or reframing the task. The goal is not to expose an exhaustive representation at all times, but to make the portions relevant to the current alignment problem inspectable and actionable.

\textbf{P2: Support Repair as a Composable Trajectory.} Our move taxonomy describes interventions at different loci because a visible symptom may have several plausible conceptual causes, and changing one element can propagate through the surrounding frame. Repair is inherently sequential in which collaborators gather evidence, form hypotheses, make interventions, observe uptake, and decide whether further repair is necessary. This perspective suggests that interfaces should support \emph{repair paths}. Early in a breakdown, when specificity is low, a useful move may primarily reveal information such as an example, contrast, probe, decomposition, or simulation may help determine what the difference actually is. Once the mismatch is localized, more direct changes to values, constraints, relations, priorities, or framing become possible. Validation can then determine whether the repair resolved the relevant difference without disturbing aspects of the interpretation that were already satisfactory.

Further, supporting repair trajectories means reducing the cost of consequential changes. A user may recognize that the system has adopted the wrong overall framing but continue making local corrections because a broader repair risks disrupting parts of the interpretation that already work. Alignment Games suggests treating repair as a \emph{composable grammar}, i.e., individual moves provide reusable operators that can be combined into larger repair sequences, adapted as new evidence emerges, and reversed or branched when a path proves unhelpful. Systems can support this compositionality through previews, reversible edits, branching alternatives, and explicit indication of which frame elements a proposed move is likely to affect. In this sense, Alignment Games is not a fixed command language, but a grammar for constructing repair trajectories at different levels of scope and abstraction. Over longer collaborations, successful frames and repair sequences could also become reusable conceptual precedents, allowing alignment to accumulate rather than be reconstructed from scratch.

\textbf{P3: Redistribute the Work of Alignment.}
Current generative systems often place most of the burden of conceptual alignment on the human. The user must recognize that a breakdown has occurred, infer what the system understood differently, decide how to communicate the correction, and determine from subsequent outputs whether the correction was taken up. When the system cannot expose or revise its interpretation, users are effectively required to adapt their own language until it produces behavior the system accepts. This resembles ``making do'' with a system's normative ground rather than collaboratively negotiating meaning~\cite{li2023beyond,subramonyam2024bridging}.

Alignment Games makes this burden visible. Conceptual repair involves at least three forms of work: \emph{diagnostic work} to determine where the mismatch lies, \emph{repair work} to formulate an intervention, and \emph{coordination work} to establish whether that intervention has been understood and incorporated. Human collaborators distribute these responsibilities dynamically and often act according to a principle of least collaborative effort~\cite{clark1986referring}. Human--AI systems need not reproduce human collaboration symmetrically, but they can be designed so that the human does not perform all three forms of work by default. Observability is one mechanism for shifting diagnostic work away from the user: the system can externalize an operational account of how it currently interprets the task. Operability can reduce repair work by making likely interventions easier to formulate. A further step is \emph{initiative}. Rather than waiting for a user to recognize a breakdown, future systems might detect uncertainty or conflicting evidence in the current frame and initiate an alignment game themselves: ``I may be interpreting `professional' as visually restrained, but your examples suggest you care more about authority and clarity. Which matters here?''

This direction reframes a capable AI collaborator as one that participates in maintaining mutual understanding. Such a system would need to track an operational representation of both its own current interpretation and relevant evidence about the user's, detect consequential differences, and select repair moves appropriate to the uncertainty and cost of the situation. This is a demanding agenda, and current systems provide only partial approximations of these abilities. The framework nevertheless gives HCI a vocabulary for asking a more precise design question: \emph{which parts of the work required for mutual understanding are currently performed by the human, and which could responsibly be supported by the system?}

\textbf{P4: Design for Task-Sufficient Alignment}
As described in Section~\ref{sec:task-sufficient-alignment}, collaborators need only align on those aspects of their situated conceptualizations that matter for the current joint activity. Thus, interfaces should support \emph{diagnostic exploration before resolution}. This may include interfaces for presenting alternatives, comparing counterfactual outcomes, preserving branches, or allowing collaborators to mark a difference as acceptable for the current task. Unexpected analogies, conceptual blends, attributes, or framings may initially appear as misalignment yet reveal possibilities that neither collaborator would have produced independently. The design objective is consequently not maximal representational convergence, but enough mutual understanding to coordinate consequential judgments and actions while preserving differences
that remain useful. In this sense, Alignment Games treats conceptual difference not only as a source of breakdown, but also as a potential resource for collaborative sensemaking and generation.

\subsection{Conceptual Alignment as Runtime Alignment}
\label{sec:runtime-alignment}
Alignment Games identifies a form of alignment complementary to dominant approaches to AI alignment. Value and instruction alignment generally concern shaping system behavior toward human preferences, intentions, or societal values~\cite{russell2019human,ouyang2022training,gabriel2020artificial}. Recent HCI work has emphasized that alignment can also occur interactively through specification, process, and evaluation during use~\cite{shen2024towards,terry2023interactive}. Work on representational, concept, and abstraction alignment meanwhile examines relationships between human and machine representations
~\cite{sucholutsky2023getting,rane2024concept,boggust2025abstraction}.

Conceptual alignment occupies a different but complementary position. Specifically, it concerns the \emph{runtime negotiation of a situated interpretation for a particular collaborative activity}. Terms such as ``clear,'' ``engaging,'' ``appropriate,'' ``professional,'' or ``important'' cannot always be fully
specified at design or training time because their task-relevant meaning depends on the immediate situation, goals, prior decisions, and collaborators involved. A system can satisfy a general instruction or preference while still construing a locally important concept differently from its user. Alignment Games consequently treats alignment not as a property achieved once in a model, but as an ongoing collaborative accomplishment. Conceptual interpretations may need to be diagnosed, negotiated, and repaired as a task
changes and new evidence becomes available. This notion of \emph{runtime alignment} complements design- and training-time approaches. Instead of attempting to anticipate every task-specific meaning in advance,
interfaces can provide mechanisms through which consequential meanings are made explicit and negotiated when they matter.

\subsection{Limitations}
\label{sec:limitations}
First, Alignment Games is motivated by theories of human situated cognition, whereas contemporary AI systems have fundamentally different architectures and may not construct or ground concepts in human-like ways. Whether AI systems possess anything appropriately described as conceptual understanding remains contested ~\cite{harnad1990symbol,searle1980minds,mollo2023vector}. We do not treat an AI's situated conceptual frame as a faithful description of its latent cognition. For AI systems, the frame is an \emph{operational interaction representation}: a task-relative account of the interpretation currently guiding behavior that can be inspected and repaired. Alignment Games likewise does not require assuming that an AI possesses human-like commitments or shared intentions. The framework can support asymmetric interaction in which
the human treats the AI primarily as a tool, as well as more bidirectional forms in which the system takes greater responsibility for maintaining alignment.

Second, conceptual alignment is difficult to observe directly. A person's situated conceptualization is only partially accessible through language, behavior, elicitation, and artifacts, while explanations generated by AI systems need not faithfully reveal their internal processing ~\cite{jacovi2020towards}. Operational frames will therefore inevitably be partial constructions rather than complete measurements of either collaborator's representation. Future work should investigate how such frames can be elicited, inferred, revised, and evaluated without requiring users to explicitly specify every aspect of their understanding. This includes lighter-weight elicitation techniques, inference from interaction histories, and methods for determining when a frame is sufficiently informative to
support repair.

Third, the alignment moves we propose are intended as a generative vocabulary, not an exhaustive catalog or empirically established policy for conceptual repair. Our examples reinterpret breakdowns reported in prior systems to show how Alignment Games could apply across domains; they do not demonstrate that the proposed repair sequences are optimal, or even that the inferred misalignment was the actual source of the reported breakdown. Empirical work is needed to examine how people select and compose alignment moves, which moves are effective under different forms of uncertainty, and whether interfaces based on Alignment Games improve diagnosis or reduce repair effort relative to conventional prompting and artifact editing. Other domains may also reveal additional moves, recurrent game structures, or different conditions for task-sufficient alignment.

Finally, making conceptual representations more observable introduces risks. Frames may externalize goals, preferences, experiences, and priorities from which increasingly detailed user models can be constructed ~\cite{shaikh2025creating}. Human-like representations of an AI's ``understanding'' may also encourage unwarranted attributions of agency or comprehension~\cite{peter2025benefits}. Systems should make clear the operational status of such representations and provide meaningful control over what conceptual information is stored, reused, or shared. Moreover, Alignment Games addresses the narrower problem of negotiating \emph{task-situated meaning}; it should not shift responsibility for broader questions of safety, fairness, cultural values, or societal alignment onto individual users.
\section{Conclusion}

Conceptual misalignment arises when collaborators construct different task-relevant interpretations of the concepts guiding their joint activity. These differences may remain hidden beneath shared language or arise because collaborators organize the situation around different concepts, attributes, relations, or priorities. We introduced \emph{Alignment Games} as a framework for making such differences visible and actionable during interaction. The framework combines a process model of situated conceptualization, a task-relative frame representation, and a composable vocabulary of alignment moves spanning the \situationterm{situation}, \processterm{processes} of conceptualization, and the resulting \frameterm{frame}. Together, these elements position conceptual alignment as a runtime collaborative process for reaching task-sufficient mutual understanding and provide a foundation for designing AI systems that help people shape the interpretations that guide generation.

\bibliographystyle{ACM-Reference-Format}
\bibliography{99_refs}

\newpage
\appendix
\section{Determinants of Conceptual Misalignments}
\label{sec:a_variations}
Here we describe the \textit{structural conditions} that make conceptual misalignment inevitable. Specifically, differences in situational cues, internal task context, prior knowledge and expertise, modes of acquisition, culture, and concept-level properties systematically alter which attributes are activated, how values are bound, and which constraints or ideals are applied. These factors continuously influence conceptualization, particularly the construction of \textit{ad hoc} concepts that are assembled to meet local task demands. Our organization of factors mirrors the conceptualization model presented in Section~\ref{sec:model}. Additionally, we discuss variance due to individual cognitive effort and noisy communication channels. 

\subsection{Variability of Conceptual Inputs}

\subsubsection{Situational Variance}
Certain aspects of the situation influence the likelihood of conceptual misalignment. The \textit{grain size} of a situation, which refers to its ``spatial and temporal extent,'' can shape what collaborators consider relevant in a situation\cite{yeh2006situated}. For instance, situations with broader grain sizes increase uncertainty about which elements or concepts should be activated; refining a single office poster for an internal event differs markedly from designing a marketing campaign poster used for months across contexts. Additionally, the \textit{tangibility}, which refers to whether the situation is real or imagined, can cause misalignment~\cite{yeh2006situated}. Real-life situations provide perceptual grounding and shared reference points, offering a basis for understanding and interpretation. In contrast, hypothetical or counterfactual scenarios lack these anchors, forcing participants to rely more heavily on inference and increasing variability in their conceptual frameworks. 

Furthermore, a situation's \textit{familiarity} and an individual's perceived \textit{psychological distance} modulate conceptualization. Familiar situations activate established knowledge and scripts~\cite{barsalou2008grounded, schank2013scripts}, reducing cognitive effort but risking rigidity through retrieval-based processing and functional fixedness~\cite{logan1988toward, dane2010reconsidering}. According to construal level theory~\cite{trope2010construal}, greater temporal, spatial, social, or hypothetical distance induces higher-level, more abstract construals, while proximity encourages concrete reasoning. Collaborators who differ in familiarity or psychological distance therefore approach the same task with \textit{distinct} abstractions, goals, or salient features, making situational variance a source of misalignment. 

\subsubsection{Internal Task Context Variance}
A person's internal task context, such as goals, motivation, and affective state (i.e., their ``internal situational elements~\cite{barsalou2018moving}''), shapes conceptual understanding. For instance, goals regulate perception and conceptual activation, as Hommel et al. note, goal-directed action requires ``the perceptual selection of certain aspects of environmental information, while other aspects are ignored or rejected~\cite{hommel2001theory}''. Barsalou's work on goal-derived categories shows that the \textit{ideality} of a concept, i.e., how closely its aspects relate to goal achievement, determines their accessibility~\cite{barsalou1983ad}, and goal-relevant attributes are more readily retrieved~\cite{hass2019idea}. Consequently, differing or underspecified goals easily yield conceptual misalignment. The \textit{structure} of goals also shapes construal, where people pursuing long-term superordinate goals (the ``why'') form more abstract representations \cite{vallacher1987people, trope2003temporal}, while those focused on immediate, subordinate goals (the ``how'') operate more concretely. When collaborators differ in goal abstraction, their frame diverges. Finally, \textit{affective states} modulate conceptualization, where positive moods broaden category inclusion and promote global processing~\cite{isen1984influence, gasper2002attending, han2014emotions}, whereas negative moods heighten focus on local, detailed features.

\subsubsection{Prior Knowledge}
Prior knowledge and experience fundamentally shape how people construct, interpret, and communicate concepts. Through \textit{concept induction}, individuals generalize from previous encounters to construct conceptual structures that guide understanding and action~\cite{tenenbaum2011grow}. Depending on the learning mechanism --- prototype similarity~\cite{posner1968genesis, rosch1975family}, exemplar similarity \cite{medin1978context}, or causal theory-building \cite{murphy1985role, gopnik1993we} --- different aspects of experience determine which features are encoded, how they are related, and which values are typical.

\textit{Differences in Concept Induction and Expertise:} The distribution and diversity of experienced examples influence the internal structure of concepts, i.e., the attributes associated with them, their organization, and the typical values. While feature frequency matters, the \textit{central tendency} of features and their \textit{diagnosticity} for achieving conceptual goals are often more important~\cite{barsalou1985ideals}. From a theory-theory perspective, concept learning involves inferring the causal relations that make attributes cohere~\cite{murphy1985role, gopnik1997words, gopnik2012reconstructing}. Different histories of experience, evidence, or interventions thus yield different conceptual representations. Social learning and enculturated instruction also shape induction. For instance,  teachers and peers emphasize contrastive or high-diagnostic features that direct learners' attention~\cite{mervis1987child, waxman1997setters, schwartz2011practicing}. \textit{Ideals} function as attractors defining the ``best'' instance of a concept~\cite{barsalou1985ideals}, but these ideals vary with culture and context~\cite{norenzayan2002cultural}. For instance, Naous et al. demonstrate how cultural norms determine which values are even conceivable --- e.g., interpreting ``drink'' as non-alcoholic after Maghrib prayer \cite{naous2023having}. Expertise further refines conceptual structure --- taxonomists, landscapers, and park professionals categorize trees differently due to distinct goals and representations \cite{medin1997categorization}. Yet deep expertise can also produce cognitive entrenchment, reducing flexibility in conceptual reappraisal \cite{dane2010reconsidering}.

\textit{Differences in Acquisition, Culture, and Memory:}  The way concepts are acquired affects their representational richness and accessibility. Concepts learned through perceptual experience evoke richer, action-oriented simulations than those learned linguistically \cite{barsalou1999perceptual, wauters2003mode}. Abstract concepts, by contrast, rely more heavily on linguistic acquisition \cite{villani2019varieties}, which makes them less tightly tied to specific experiences. Age of acquisition also matters: early-learned concepts are retrieved more quickly and tend to be more stable, whereas later-acquired concepts remain more flexible \cite{juhasz2005age, dane2010reconsidering}. Socially mediated and enculturated learning produces narrower, more aligned conceptual categories, as seen when taxonomists share highly structured conceptual representations compared to park professionals' more pragmatic ones~\cite{medin1997categorization}. The \textit{Words as Tools} hypothesis suggests that socially transmitted concepts rely on linguistic scaffolding and shared norms, increasing coherence within but not across communities~\cite{borghi2009words}. Cultural variation further amplifies these effects, producing systematic differences in conceptual systems, category coherence, and ideals across groups~\cite{malt1995category, kovecses2005metaphor, atran2008native, waxman2007folkbiological, ojalehto2015perspectives, barsalou2023implications}.

\textit{Differences in Prior Knowledge structures and Memory:} 
Beyond structured conceptual representations, individuals differ in the episodic and sensory experiences that feed conceptual activation. Barsalou describes these as ``populations of situated conceptualizations'' unique to each person, making large individual differences in grounded cognition ``the rule, not the exception~\cite{barsalou2020challenges}''. The organization of semantic networks also varies: people with more flexible or creative reasoning often exhibit looser, more associative conceptual structures, supporting divergent thought and reinterpretation \cite{kenett2019semantic}. Such variation in the content, acquisition, and organization of prior knowledge profoundly shapes how collaborators represent and align concepts, making prior knowledge a deep and persistent source of misalignment.

\subsection{Conceptual Sources of Variance}

\subsubsection{Frame/Concept Variance}
Concepts, both long-acquired and ad hoc, exhibit wide variability in structure and use. This variability shapes how collaborators interpret and apply the same terms, making \textit{concept-level variance} a key driver of misalignment. A classical dimension is the continuum from \textit{concrete} to \textit{abstract} concepts. Concrete concepts (e.g., ``door,'' ``chart'') refer to perceptible, spatially bounded entities, while abstract concepts (e.g., ``love,'' ``beauty'') often refer to internal, relational, or temporally extended phenomena~\cite{barsalou2018moving}. Concrete concepts tend to elicit stronger perceptual grounding and less variance in associated attributes~\cite{borghi2014words}, whereas abstract concepts rely more on social and introspective contexts and exhibit broader variability~\cite{barsalou2018moving, borghi2017challenge}. Dual Coding Theory~\cite{paivio1990mental} and related accounts of perceptual strength suggest that concrete concepts are more imageable and tied to sensory experience, supporting richer perceptual simulation.

A concept's \textit{level of abstraction} can also shift dynamically depending on how it is construed. Construal Level Theory captures this flexibility: representing a ``cellular phone'' as ``a communication device'' omits detail while increasing generality~\cite{trope2010construal}. Other frameworks describe related dimensions, such as \textit{context availability} --- how easily a concept evokes a situation~\cite{schwanenflugel1983differential} --- and \textit{situational systematicity}, the degree to which a concept is consistently embedded in similar contexts~\cite{davis2020situational}. Concepts with low situational systematicity (e.g., ``love'') activate diffuse or variable contexts, whereas those with high systematicity (e.g., ``spinach'') activate narrow, stable ones. Even abstract concepts, however, often recruit situated imagery, typically social or introspective in nature \cite{barsalou2005situating}. The related notion of \textit{semantic diversity} captures the breadth of contexts in which a concept participates~\cite {hoffman2016meaning}, with highly diverse concepts supporting flexible but potentially ambiguous use.

Beyond contextual flexibility, concepts also differ in their positioning within larger semantic networks. Concepts with many associative connections (i.e., high degree) are retrieved more easily and can facilitate lexical decisions~\cite{buchanan2001characterizing, pexman2007neural}, yet dense connectivity can also diffuse activation, producing interference -- the fan effect \cite{anderson1999fan}. These network structures differ for concrete and abstract concepts: concrete concepts cluster within dense local subnetworks that connect strongly to a few contexts, while abstract concepts form sparser links across many contexts~\cite{lakhzoum2021semantic, paivio1968concreteness}. Moreover, associative links in semantic memory are directionally asymmetric, affecting activation flow between concepts~\cite{popov2019semantic}.

Finally, concepts vary in \textit{semantic richness} --- the number and diversity of attributes they encode \cite{recchia2012semantic}. Rich concepts support flexible and metaphorical use across domains. They also differ in their \textit{hierarchical position}, abstract concepts typically occupy higher levels in conceptual hierarchies and encompass broader generalizations~\cite{barsalou2005situating}. Expertise modifies this structure, i.e., experts access subordinate distinctions as readily as novices access basic-level categories~\cite{tanaka1991object} and organize knowledge by underlying principles rather than surface features~\cite{chi1981categorization}. Thus, variability in abstraction, contextual embedding, associative structure, and semantic richness together determine the interpretive latitude of a concept, making concept-level variance a persistent source of conceptual misalignment in collaboration.

\subsubsection{Attribute Variance}
There is substantial variability in the attributes people associate with a concept. Even for concrete concepts, Barsalou found only moderate agreement across individuals (correlations of $0.3$ – $0.6$) on which attributes were relevant~\cite{barsalou1981instability}, noting that this sample was unusually homogeneous. Within individuals, attribute retrieval is also unstable --- showing only $0.80$ test–retest reliability over two weeks, meaning roughly 40\% non-overlap~\cite{barsalou1993linguistic}. These findings suggest that a concept’s attribute structure is not fixed but fluid, varying both across and within people over time. A key determinant of attribute inclusion is central tendency, i.e., how closely an attribute aligns with a concept's prototypical representation~\cite{rosch1975family}. For natural categories, people rely on statistical regularities, yet this alone cannot account for conceptual variance. Theories about how features cohere play a complementary role: as Murphy notes, \textit{``similarity may be a by-product of conceptual coherence rather than its determinant~\cite{murphy1985role}''}. Attributes are thus selected not just because they are frequent but because they make sense within a conceptual or causal framework.

For goal-derived categories, \textit{ideals} rather than averages determine which attributes matter \cite{barsalou1985ideals}. An ``ideal lunch,'' for example, might emphasize speed or calorie content depending on one's goals, shaping which attributes are foregrounded. Likewise, causal and constraint relations between attributes influence how people structure a concept: ``the premises are often phrased as conditional relationships,'' reflecting people's implicit understanding of why properties cohere as they do~\cite{Johnson2000explanatory}. Such causal and normative differences yield divergent attribute sets even when collaborators appear to be referring to the same concept.

\subsubsection{Value Variance}
People often instantiate values for shared attributes in quite different ways. These differences arise from default expectations, contextual reasoning about ideals or constraints, and divergent optimization priorities.
A common source of value variance comes from \textit{default} or \textit{typical} values. In prototype theory, concepts are organized around central tendencies, i.e., based on the average or most frequent value across exemplars \cite{rosch1975family}. For example, the attribute ``height'' for the concept ``chair'' might default to approximately 18 inches. Such defaults reflect statistical learning from prior experience but vary across communities depending on the distributions of examples people encounter. Misalignment occurs when collaborators assume that their defaults are shared when, in fact, they are not~\cite{barsalou1981instability}.
Variance also appears in the breadth of acceptable ranges. One person may treat ``a reasonable lunch price'' as between \$8 and \$12, while another considers anything from \$5 to \$30 acceptable. These differences often reflect prior exposure, socioeconomic context, or cultural norms~\cite{markus1991culture, medin2004native}. Broader ranges promote flexibility and inclusivity but can increase the risk of misalignment when precision is required.

A further source of divergence lies in \textit{optimization} trade-offs. When multiple attributes compete, people vary in which compromises they are willing to make~\cite{keeney1993decisions, payne1993adaptive}. In design, for instance, one collaborator may prioritize speed of production, while another emphasizes durability, resulting in different judgments of what constitutes ``good enough.'' Ultimately, values are embedded in broader, normative, and constraint-based systems. For some, acceptable values are bounded by functional or physical constraints; for others, they are shaped by ethical or cultural commitments such as sustainability, fairness, or equity. These normative orientations often remain implicit, making them especially prone to breakdowns in conceptual alignment.

\subsection{Process Variance}

\subsubsection{Perceptual and Attentional Variance}
Collaborators differ in their perceptual abilities and attentional focus, leading them to extract different information from the same environment. From a \textit{signal detection} perspective, some individuals are more sensitive to particular features, affecting which aspects of a situation become available for conceptual activation. These perceptual asymmetries shape what information is encoded and, consequently, which concepts or frames are likely to be invoked. From a \textit{top-down} perspective, perception is strongly guided by prior knowledge, experience, and task goals. Expert chess players, for instance, can perceptually ``chunk'' the board into meaningful configurations, allowing them to recognize strategic patterns more efficiently~\cite{chase1973perception}. 

Similarly, differences in prior expertise or goals shape which features of a situation appear salient, influencing how collaborators construe the same environment. Attention further amplifies these perceptual differences. Individuals may focus on distinct aspects of a shared task, leading to divergent situational framings. Joint attention --- the ability to coordinate focus on a shared referent --- is crucial for successful collaboration~\cite{brennan1995centering} and learning~\cite{tomasello1986joint}. However, maintaining joint attention can be disrupted under high attentional load~\cite{swallow2013attentional}. Visual cues such as gaze can help mitigate these breakdowns by signaling focus and relevance even under demanding perceptual conditions~\cite{xu2011gaze}.

\subsubsection{Retrieval variance}
Retrieval variance is closely related to the aspects previously discussed in prior knowledge. However, key differences include \textit{frequency} and \textit{recency} of concept use \cite{murphy2004big}, the contextual/situational \textit{relevance} of those concepts, the \textit{strength} of initial encoding, and the degree of \textit{interference} from competing information~\cite{anderson1999fan}. 

\subsubsection{Reasoning Variance}
Differences in reasoning strategies for frame induction and value setting can lead collaborators to construct markedly different conceptual understandings.
In analogical reasoning, for example, Hummel and Holyoak show that ``people will produce different, internally consistent mappings for the same analogy,'' and that such mappings are sensitive to the order in which information is processed~\cite{hummel1997distributed}. Similarly, Thibodeau and Boroditsky demonstrate that the timing of a metaphor or analogy influences how people reason metaphorically~\cite{thibodeau2011metaphors}. Expertise also shapes analogical reasoning. Experts tend to prefer medium-range analogies over near or far ones in design contexts~\cite{kalogerakis2010developing} and are more adept at identifying deep structural relations rather than surface similarities~\cite{ozkan2013cognitive, chi1981categorization}. Other forms of conceptual reasoning, such as blending and conceptual combination, are likewise influenced by both conceptual structure and reasoning process. For instance, conceptual combination relies less on superficial feature matching and more on causal relations among features~\cite{Johnson2000explanatory}. Thus, differences in causal understanding can interact with reasoning strategies, producing distinct conceptual outcomes.

Reasoning also varies across individuals in cognitive architecture and representational structure. Variability in the topology of semantic networks affects conceptual flexibility and reasoning capacity~\cite{kenett2019semantic}. Individual differences in conceptual reasoning ability have been empirically documented~\cite{hammer2019individual}, and reasoning performance depends on working-memory capacity, attention control, and secondary memory~\cite{unsworth2014working}. Expertise further enhances the ability to form and manipulate abstract representations and to organize problems around deep structural principles~\cite{chi1981categorization}.
Finally, meta-reasoning --- the monitoring and regulation of one's own reasoning processes --- introduces additional variance. Differences in metacognitive control and confidence calibration~\cite{ackerman2017meta, buchel2013metacognitive} can influence how people interpret analogies or evaluate their own understanding. Perceived fluency, for instance, can increase subjective certainty even when actual comprehension is limited~\cite{topolinski2010immediate}. Together, these forms of reasoning vary strategically, structurally, and metacognitively, highlighting how collaborators may pursue different inferential paths toward what appear to be shared concepts, leading to subtle but consequential misalignments.

\subsection{Asymmetry of thinking between agents}
Even when collaborators share the same information, they may differ in the depth and extent of reasoning devoted to a concept or situation. At any given point in a collaboration, one person may have engaged in more sustained, goal-directed reasoning, while another relies on shallower or more heuristic processing. Such asymmetries affect problem solving and coordination: greater deliberate reasoning generally improves performance on complex problems \cite{kramer2023testing}, whereas under time pressure or limited expertise, people often operate with cognitively underspecified models --- partial, heuristic representations that substitute for full mental models~\cite{reason1992cognitive}. These dynamics parallel Kahneman's distinction between System 1 and System 2 thinking \cite{kahneman2011thinking}. When one collaborator is reasoning analytically while another depends on fast, intuitive judgments, misalignments can emerge not from differing beliefs but from differences in how much or how deeply each has reasoned about the problem. 

\subsection{Communication: Compression, Specificity, and Noise}
When collaborators communicate, they must express rich conceptual representations through limited linguistic channels. Language serves as a form of \textit{compression}, i.e., we cannot externalize the full content of our internal frames, so we use words and other symbols as pointers to shared concepts. This compression is necessarily \textit{lossy} in which meanings are approximated, not fully transmitted. Following Grice's maxim of quantity~\cite{grice1975logic}, speakers aim to be informative without being redundant, striking a balance between communicative efficiency and clarity. However, this economy of expression often leads to \textit{underspecification} and \textit{ambiguity}, requiring interlocutors to infer the intended meaning.
Ambiguity arises when utterances admit multiple interpretations. 

\textit{Polysemy}, where a word has several distinct senses, is a significant source of such uncertainty~\cite{klein2001representation}. An example of this is the widely shared AI error of it describing the difference between sauce and dressing: ``The main difference between a sauce and a dressing is their purpose: sauces add flavor and texture to dishes, while dressings are used to protect wounds''~\cite{msn_ai_plugin_search}. Here, the AI system encounters a polysemous word, ``dressing", and activates the wrong concept; it chooses wound dressings as opposed to dressings in the context of food (dressing as a condiment ). Humans face similar challenges when contextual cues are insufficient to disambiguate meaning. Concrete terms can often be grounded in shared perceptual referents, but abstract words --- those referring to internal states, relationships, or temporally extended phenomena --- lack direct anchors and therefore exhibit higher interpretive variability.

Underspecification further compounds misalignment. As Clark notes, speakers routinely omit detail, trusting that listeners will fill in gaps using shared common ground \cite{clark1996using}. This pragmatic economy assumes overlap in prior knowledge and situational awareness; when that overlap is weak, listeners’ inferences diverge, producing conceptual drift. Finally, vagueness which refers to a deliberate lack of precision, also shapes alignment. Austin observed that ```Vague' is itself vague''~\cite{austin1975things}, yet vague language serves several communicative functions: it can reflect genuine uncertainty, reduce cognitive effort, or signal the relative unimportance of a detail~\cite{jucker2003interactive}. Vagueness can also be strategic in directing attention toward salient features or conveying meta-information such as confidence or relevance. However, when speakers and listeners interpret vagueness differently, it introduces noise into the alignment process, obscuring which conceptual elements are meant to be shared.


\end{document}